\documentclass[conference,hidelinks]{IEEEtran}%,10pt
\usepackage{amsmath,graphicx}
\usepackage{url}
\usepackage{amsfonts}
\usepackage{amssymb}
\usepackage{amsbsy}
\usepackage{bm}
\usepackage{booktabs}
\usepackage{times}
\usepackage{color}
\usepackage[noadjust]{cite}
\usepackage{enumerate}
\usepackage{enumitem}
\usepackage{tikz}
\usepackage{textcomp}
\usepackage{graphicx}
\usepackage{amsthm}
\usepackage{dsfont}
\usepackage{tabularx}
\usepackage{placeins}
\usepackage[para]{threeparttable}
\usepackage{multirow}
\usepackage{multicol}
\usepackage[binary-units]{siunitx}
\usepackage{subfigure}

\usepackage[noadjust]{cite}

\usepackage{footnote}
\makesavenoteenv{tabular}
\makesavenoteenv{table}
\usepackage{soul}

\usepackage{pgfplots}
\pgfplotsset{compat=1.14}

 \usepackage[colorlinks=true,urlcolor=blue]{hyperref}

\newcommand{\subparagraph}{}
\usepackage[compact]{titlesec}

\titleformat{\paragraph}[runin]
{\normalfont\normalsize\itshape}{\theparagraph}{1em}{}

\titlespacing*{\section} {0pt}{1.5ex plus 1ex minus .2ex}{0.8ex plus .2ex}
\titlespacing*{\subsection} {0pt}{1.2ex plus 1ex minus .2ex}{0.8ex plus .2ex}
\titlespacing*{\subsubsection}{0pt}{1ex plus 1ex minus .2ex}{0.8ex plus .2ex}
\titlespacing*{\paragraph} {0pt}{1ex plus 1ex minus .2ex}{1em}

\newcolumntype{C}{>{\centering\arraybackslash}X}
\newcolumntype{R}{>{\raggedleft\arraybackslash}X}
\newcolumntype{x}[1]{>{\centering\arraybackslash\hspace{0pt}}p{#1}}

\newcommand{\secref}[1]{{Sec.~\ref{#1}}}

\newcommand{\figref}[1]{{Fig.~\ref{#1}}}
\newcommand{\tabref}[1]{{Tbl.~\ref{#1}}}

\title{On the Feasibility of FPGA Acceleration of Molecular Dynamics Simulations\\
\vspace{0.3cm}
\Large Technical Report (v0.1)\\
\vspace{-0.35cm}
}
\author{
\IEEEauthorblockN{Michael Schaffner$^\dag$, Luca Benini$^\dag$}
\IEEEauthorblockA{$^\dag$ETH Zurich, Integrated Systems Lab IIS, Zurich, Switzerland\\\vspace{-1.2cm}}
}
\begin{document}
% automatic et al abbreviations
\bstctlcite{IEEEexample:BSTcontrol}
%
% make the title area
\maketitle

\begin{abstract}
Classical molecular dynamics (MD) simulations are important tools in life and material sciences since they allow studying chemical and biological processes in detail. However, the inherent scalability problem of particle-particle interactions and the sequential dependency of subsequent time steps render MD computationally intensive and difficult to scale. To this end, specialized FPGA-based accelerators have been repeatedly proposed to ameliorate this problem. However, to date none of the leading MD simulation packages fully support FPGA acceleration and a direct comparison of GPU versus FPGA accelerated codes has remained elusive so far.

With this report, we aim at clarifying this issue by comparing measured application performance on GPU-dense compute nodes with performance and cost estimates of a FPGA-based single-node system. Our results show that an FPGA-based system can indeed outperform a similarly configured GPU-based system, but the overall application-level speedup remains in the order of 2$\times$ due to software overheads on the host. Considering the price for GPU and FPGA solutions, we observe that GPU-based solutions provide the better cost/performance tradeoff, and hence pure FPGA-based solutions are likely not going to be commercially viable. However, we also note that scaled multi-node systems could potentially benefit from a hybrid composition, where GPUs are used for compute intensive parts and FPGAs for latency and communication sensitive tasks.
\end{abstract}

\section{Introduction}
Classical molecular dynamics (MD) simulations \cite{Karplus1990} are important tools in life and material sciences since they allow studying chemical and biological processes in detail. For example, this enables researchers to study drug-target bindings for drug discovery purposes \cite{DeVivo2016} or to analyze protein folding processes to understand their biological function \cite{Shaw2009}.

However, MD is computationally intensive and difficult to scale due to the sequential dependency between subsequent timesteps and the many particle-particle interactions. The timescales of interest are often orders of magnitude larger than the simulation time steps (i.e., \SI{}{\nano\second} or \SI{}{\micro\second} timescales versus \SI{}{\femto\second} time steps), which results in long simulation times even on HPC computing infrastructure.

To this end, several approaches have been pursued to improve simulation performance, ranging from novel algorithms to approximate forces between particles \cite{Hardy2016multilevel} over algorithmic tweaks \cite{Krieger2015} and biasing methods such as enhanced sampling methods \cite{Spitaleri2018}, to custom hardware solutions, such as the MDGRAPE systems from Riken \cite{Ohmura2014} and the Anton-1/2 supercomputers developed by D.E. Shaw Research LLC \cite{Shaw2008,Shaw2009,Shaw2014}. However, we observe that algorithmic improvements are often very problem specific, and it often takes a long time until they are adopted by major production software packages. Hence, the core algorithms used in classical MD simulations have largely remained in recent years, and most simulation speed improvements are due to the use of MPI parallelization in combination with GPGPUs that handle the computationally dominant parts. Specialized supercomputers such as the Anton systems are very inaccessible and expensive, and are hence not widely used today.

Besides these MPI and GPU-based solutions, FPGA accelerators have repeatedly been proposed as a viable alternative to accelerate the compute intensive parts \cite{malicta2017molecular,azizi2004reconfigurable,Escobar2016,Khan2013fpga,Chiu2010towards,Sanaullah2016gpu,Sanaullah2016fpga,Sheng2014design,Humphries20143d,Humphries20133d,Chiu2011efficient,Chiu2009efficient,Kasap2012}
. However, these studies only show estimated or measured speedup with respect to older CPU implementations. To date none of the leading MD simulation packages fully support FPGA acceleration \cite{Cong2016} and a direct comparison of GPU versus FPGA accelerated codes has remained elusive so far.

This report aims at shedding some light onto the questions whether and how FPGAs could be used to accelerate classical MD simulations in the scope of biochemistry, and whether such a solution would be commercially viable. To this end, we revisit existing FPGA architectures, model their behavior on current FPGA technology and estimate the performance and price of an FPGA accelerated system in order to compare with GPU accelerated solutions. We focus on single node systems in this report (possibly carrying several accelerator cards) since these represent the most common configuration employed today\footnotemark. Typical MD problems with in the order of 100k atoms do not scale well across several nodes, and hence it is most economic to run these simulations on accelerator-dense single node systems.

Our results show that, in principle, FPGAs can be used to accelerate MD, and we estimate full application-level speedups in the order of 2$\times$ with respect to GPU-based solutions. However, our estimates also indicate that this speedup is likely not high enough to compensate for the increased cost and reduced flexibility of FPGA-based solutions. Hence we conclude that FPGAs are likely not well suited as a replacement for GPU accelerators. However, we observe that other aspects of FPGAs like the low-latency networking capabilities could be leveraged to augment scaled multi-node systems and ameliorate scalability issues by providing network-compute capabilities in addition to GPU acceleration.

This report is structured in three main sections: \secref{sec:relwork} summarizes background and related work, performance benchmarks of two widely used software packages are given in \secref{sec:bench}, and in \secref{sec:perfEst} we provide the FPGA estimates and comparisons with GPU-based systems.
\section{Background and Related Work}
\label{sec:relwork}
%
%This section gives a brief overview of MD and existing work on hardware acceleration.
%
\footnotetext{\url{http://ambermd.org/gpus/recommended_hardware.htm}}
\subsection{Classical MD}
This section gives a brief overview of MD, for more details we refer to \cite{Karplus1990,rapaport2004art,DeVivo2016,hansson2002molecular}.
\subsubsection{Simulation Loop and Force-Fields}
A typical biomolecular MD simulation consists of a macromolecule that is immersed in a solvent (e.g., water). Each atom in the system is assigned a coordinate $\mathbf{x}_i$, velocity $\mathbf{v}_i$ and acceleration $\mathbf{a}_i$. The aim of MD is to simulate the individual trajectories of the atoms, and this is done by integrating the forces acting on each particle. The integration is carried out in discrete timesteps, with a resolution in the order of \SI{2}{\femto\second}, and the forces are calculated by taking the negative derivative of a potential function $V$ w.r.t. to each particle coordinate $\mathbf{x}_i$:
\begin{equation}
\mathbf{f}_i\left(t\right)=m_i\mathbf{a}_i\left(t\right)=-\frac{\partial V\left(\mathbf{x}\left(t\right)\right)}{\partial (\mathbf{x}_i\left(t\right)}
\end{equation}
where the potential function is based on classical Newtonian mechanics. The potential typically comprises the following terms:
\begin{equation}
\begin{aligned}
\label{eq:ff}
V&=\sum_i^{\text{bonds}}\frac{k_{l,i}}{2}\left(l_i-l_{0,i}\right)^2
+\sum_i^{\text{angles}}\frac{k_{\alpha,i}}{2}\left(\alpha_i-\alpha_{0,i}\right)^2\\
&+\sum_i^{\text{torsions}}\left(\sum_k^{M}\frac{V_{ik}}{2}\cos\left(n_{ik}\cdot\theta_{ik}-\theta_{0,ik}\right)\right)\\
&+\sum_{i,j}^{\text{pairs}}\epsilon_{ij}\left(\left(\frac{r_{0,ij}}{r_{ij}}\right)^{12}-2\left(\frac{r_{0,ij}}{r_{ij}}\right)^6\right)\\
&+\sum_{i,j}^{\text{pairs}}\frac{q_i q_j}{4 \pi \epsilon_0 \epsilon_r r_{ij}}
\end{aligned}
\end{equation}
Such a set of functions and the corresponding parameterization is often also called a \emph{force-field} (examples for specific force field instances are CHARMM36m, GROMOS36 or AMBERff99). The first three terms in \eqref{eq:ff} cover the interactions due to bonding relationships and are hence also called the bonded terms. The last two terms cover the van der Waals and electrostatic (Coulomb) interactions, and are hence called the non-bonded terms. We can already anticipate from these equations that the non-bonded terms will play a dominant role during force calculation since it is a particle-particle (PP) problem with $\mathcal{O}(N^2)$. The bonded terms are usually much less intensive ($\mathcal{O}(N)$) since we do not have bonds among all particles in the system.

Most simulations are carried out using periodic boundary conditions (PBC) in order to correctly simulate the bulk properties of the solvent. This means that the simulation cell containing the $N$-particle system above is repeated infinitely many times in all directions, which has implications on the algorithms used for the non-bonded forces.

A typical MD simulation has an outer loop that advances the simulation time in discretized steps in the order of \SI{}{\femto\second}, and in each time step, all forces in the system have to be calculated, the equations of motion have to be integrated, and additional constraints\footnote{Constraints are usually used to eliminate high-frequency oscillations of H atoms that would lead to integration issues otherwise, see also \cite{Hess1997,Hess2008,Miyamoto1992}.} have to be applied before restarting the loop.
\subsubsection{Evaluation of Non-bonded Interactions}
The first term with $1/r^12$ and $1/r^6$ modelling the Van der Waals interaction decays very quickly. Therefore, a cutoff radius r can be used (usually in the order of 8-16 Angstroms), and only neighboring particles within the spanned sphere are considered during the evaluation of this term.

The Coulomb interaction on the other hand is more difficult to handle, as it decays relatively slowly ($1/r$). Earlier methods also just used a cutoff radius, but this may introduce significant errors. To this end, different algorithms that leverage the periodic arrangement have been introduced. The most prominent one is the Ewald sum method (and variations thereof). This is basically a mathematical trick to split the Coulomb interaction into two separate terms: a localized term that is evaluated in space-domain, and a non-local term that is evaluated in the inverse space (i.e., in the spatial frequency domain / k-space). This decomposition enables efficient approximation methods that have computational complexity $\mathcal{O}(N \log(N))$ instead of $\mathcal{O}(N^2)$. One such method that is widely used together with PBC today is called \emph{Particle-Mesh Ewald} (PME) \cite{esselink1995comparison,petersen1995accuracy,Essmann1995,Shan2005}. Basically, the spatial term is handled locally with a cutoff radius (same as for the Van der Waals term above), and the second is approximated using a gridded interpolation method combined with 3D FFTs (hence the $\mathcal{O}(N \log(N))$) complexity). The spatial term is sometimes also called the \emph{short-range} interaction part, whereas the inverse space term is called the \emph{long-range} interaction part.

In terms of computation and communication, the non-bonded terms are the heavy ones, and consist of two cutoff-radius limited PP problems, and two 3D FFT problems. As we shall see later in \secref{sec:bench}, these non-bonded forces account for more than 90\% of the compute time in an unparallelized MD simulation. Moreover, an intrinsic problem of MD is that the simulation time steps are sequentially dependent, which inherently limits the parallelizability, and means that we have to parallelize a single time step as good as possible (which essentially creates a communication problem). Note that the dataset sizes are very manageable and only in the order of a couple of MBytes.
\subsubsection{Other Approaches}
Today, the most serious contenders for replacing PME electrostatics seem to be fast multipole methods (FMM) \cite{Arnold2013,greengard1987fast,white1994derivation,esselink1995comparison,challacombe1997periodic}, multi-level summation methods (MSM) \cite{Hardy2015msm,Hardy2016multilevel} and multi-grid (MG) methods \cite{Skeel2002}:
\begin{itemize}
\item\textbf{FMM:} The FMM makes use of a hierarchical tree decomposition of simulation space and interacts particles with multipole approximations of the farfield, thereby giving rise to $\mathcal{O}\left(N\right)$ complexity. For low order expansions, the multipole approximation of the electrostatic potential function has large discontinuities. This leads to drifts of the total energy over time, and this violation of energy conservation is not acceptable in MD applications. To resolve this issue, one has to resort to high order expansions that are more accurate, but also more complex to calculate. With sufficiently high expansion orders, the FMM can be slower \cite{petersen1995accuracy,Hardy2016multilevel} than fast PME algorithms in the important range of $N$ = 10k - 100k particles (despite the fact that FMM has better asymptotic complexity than PME)\footnote{See also these LAMMPS and GROMACS forum entries: \url{http://lammps.sandia.gov/threads/msg72001.html}, \url{http://www.gromacs.org/Developer_Zone/Programming_Guide/Fast_multipole}}.
\item\textbf{MG:} These methods discretize the Laplace operator and solve the resulting linear system with a multigrid solver, which results in linear complexity as well. however, these methods suffer from relatively large discretization errors compared with FFT methods, thereby leading to clear energy drifts that need to be corrected for \cite{Arnold2013}.
\item\textbf{MSM:} This method splits the interaction kernels smoothly into a sum of partial kernels of increasing range and decreasing variability, and the long-range parts are interpolated from grids of increasing coarseness, and hence MSM provides $\mathcal{O}\left(N\right)$ complexity, too. The MSM is relatively new, and hence not widely adopted yet. It is unclear how it compares to PME in speed, but it is shown to provide better performance than FMM in the case of highly parallel computation and it can be used for nonperiodic systems, where FFT-based methods do not apply \cite{Hardy2016multilevel}.
\end{itemize}
Due to the reasons mentioned, FMM, MSM and MG methods have not yet been widely adopted by common software packages. Another reason is that in a direct comparison \cite{Arnold2013}, FFT-based methods are still among the most efficient in performance and stability -- and hence it is difficult to motivate a migration to an new experimental algorithm if it not absolutely required. The $\log\left(N\right)$ contribution of FFT based methods does not seem to be visible for common system sizes. Hence, for single-node systems not operating at scale, FFT-based methods still seem to be the best algorithmic choice for simulations with PBC due to their efficiency and algorithmic simplicity. As we will see later in Sections~\ref{sec:bench} and \ref{sec:perfEst}, one way to address the communication bottleneck within PME can be addressed by dedicating it to a single FPGA or GPU, where it can be executed at very high speeds.
\subsubsection{Classical versus Ab-Initio MD}
Note that classical MD should not to be confused with ab-initio MD (AIMD) that operates on quantum-mechanical potential approximations. While AIMD has the advantages of being more accurate and not requiring explicit parameterization, it is also computationally much more involved and hence limited to small MD systems comprising a few 100 to 1000 particles. Classical MD works on a coarser abstraction\footnote{Mixed simulation modes where some potential calculations of classical MD are augmented with QM are possible, but not considered in this report.} by employing classical mechanics in the form of force-fields that can be evaluated more efficiently, hence allowing to simulate larger systems with up to several 100k atoms. Although classical MD simluations are less accurate than their AIMD counterparts, they are often accurate enough for many biomolecular systems, where AIMD would not be feasible due to amount of particles involved. The drawback of classical MD is that the force-fields need to be parameterized with experimental data, and they can often not model chemical reactions as the bonds are static.

An interesting new algorithmic development for AIMD-based approaches is to use neural networks (NN) to accelerate the evaluation of the quantum-mechanical potentials \cite{Smith2017}.
%
%%%%%%%%%%%%%%%%%%%%%%%%%%%%%%%%%%%%%%%%%%%%%%%%%%%%%%%%%%%%%%
%ASIC literature by Anton, MD GRAPE etc
\subsection{Dedicated ASICs for classical MD}
There exist only a few dedicated systems that have been built to enhance MD simulation performance. Two older ones are the MD-ENGINE \cite{toyoda1999development} and MDGRAPE-3 \cite{Taiji2003}. The MD-ENGINEis a simple ASIC coprocessor that evaluates non-bonded forces only (with a non-optimized Ewald sum method which is $\mathcal{O}\left(N^2\right)$). Several such accelerators can be attached to a SPARC host system that carries out the rest of the calculations. Compared to newer architectures, the system does not scale well since it is still based on the a direct Ewald sum method. the interpolation units use a combination of fixed point and extended single-precision (\SI{40}{\bit}) FP formats. MDGRAPE-3 is another MD co-processor chip which is similar to the MD-ENGINE above in the sense that only the non-bonded portion in the MD force calculation is accelerated, and they still use the direct Ewald sum method, which is accurate, but $\mathcal{O}(N^2)$. However, the complete system is much larger than previous ones: the complete system comprises 256 compute nodes equipped with 2 Itanium class CPUs and two MDGRAPE boards with 24 chips each. This results in a cluster with 512 CPUs and 6144 special purpose MDGRAPE chips. Also, they use a ring interconnect topology on each MDGRAPE board, which facilitates reduction operations such as force summation on one board.

Apart from these older instances above, there are also a few more recent incarnations of such special purpose machines that leverage complete SoC integration to accelerate MD in a more holistic fashion. The most notable machines are the Anton-1 \cite{shaw2008anton,Shaw2009,grossman2008simulation} and Anton-2 \cite{Shaw2014} computers by D.E. Shaw Research LLC. The MDGRAPE-4 chip from Riken \cite{Ohmura2014} is another special purpose SoC for MD that has similarities with the Anton chips, but the project does not seem to be as successful as Anton.

Both Anton generations are similar from a conceptual viewpoint, so we will mainly refer to numbers from Anton-2 in the following. The system configuration comprises a set of 512 compute nodes, interconnected with a 8$\times$8$\times$8 3D Torus, and each compute node consists of one Anton-2 ASIC fabricated in \SI{40}{\nano\metre}. The 3D torus arrangement allows for natural problem mapping using a domain decomposition and facilitates communication in all directions. The Anton chips themselves are quite innovative with respect to the previous MD accelerators in the sense that they do not only contain fixed-function accelerators, but both C++ programmable Tensilica processors and specialized \emph{pairwise point interaction modules} (PPIMs). Each Anton-2 chip contains 64 simple \SI{32}{\bit} general purpose processors with 4 SIMD lanes and two \emph{high-throughput interaction subsystem} (HTIS) tiles containing an array of 38 PPIM streaming accelerators each. The heterogenous arrangement allows Anton to perform complete MD simulations without having to closely interact with a host system. I.e., it can also accelerate the other parts of the MD simulation (the remaining 10\% of the computations), thus enabling better scalability. Moreover, Anton uses optimized PME algorithms to circumvent the $\mathcal{O}(N^2)$ issue of brute-force PP interactions. \cite{young200932x32x32} provides a detailed study of distributed 32$\times$32$\times$32 and 64$\times$64$\times$64 3D FFTs on Anton machines.

Interestingly, newer firmware versions running on Anton use real-space convolution instead of 3D FFTs as this requires less all-to-all communication phases that impact scalability. Another interesting aspect to note is that external DRAM -- albeit supported by the Anton-2 chips -- is not required even in for large scale simulations, since the complete state of the problem fits into a couple of MBytes, and hence into the distributed on-chip SRAM available on the chips.
%
%%%%%%%%%%%%%%%%%%%%%%%%%%%%%%%%%%%%%%%%%%%%%%%%%%%%%%%%%%%%%%
% FPGA story
\subsection{MD Acceleration using FPGAs}
Most studies on FPGA accelerated MD target the PP calculations \cite{Kasap2012,Chiu2010towards,Khan2013fpga,azizi2004reconfigurable,malicta2017molecular,Chiu2009efficient,Chiu2011efficient}, since they make up for over 90\% of the sequential runtime in typical simulations. Apart from these studies, only relatively few papers cover PME and other aspects \cite{Humphries20133d,Humphries20143d,Sanaullah2016fpga,xiong2017bonded}.

We note that most work in the area has been done by the group by M. Herbordt at Boston University, including the design of several variations of PP interaction pipelines \cite{Khan2013fpga,Chiu2010towards,Chiu2011efficient,Chiu2009efficient,chiu2010molecular} and a prototype where such a PP interaction pipeline is integrated into a simplified variant of NAMD \cite{Chiu2010towards}.
Their PP interaction pipelines makes use of \emph{particle filters} that build the neighborlists on-the-fly. In contrast to software-based implementations, such a filter-based approach does not require large neighborlist buffers and can be efficiently implemented using systolic filter arrays and reduced arithmetic precision in hardware (the Anton PPIMs employ a similar approach). Their preferred PP design employs 1st order piece-wise polynomial interpolation with around 6 segments and 256 LUT entries per segment \cite{Chiu2011efficient}. Since their prototype system uses relatively dated FPGA technology (Stratix III), it is difficult make comparisons with current GPU-based solutions, and updated performance estimates are required (no rigorous performance comparison with contemporary GPU technology has been carried out in their study). In \cite{Humphries20133d,Humphries20143d}, it is shown that commonly used 3D FFTs in the order of $64^3$ can be conveniently solved on single FPGAs at competitive speeds in the range of a few \SI{100}{\micro\second}. In \cite{Sanaullah2016fpga} they present an interpolator design able to carry out the charge spreading and force gathering stems within PME, and in \cite{xiong2017bonded} a preliminary analysis of bonded-force computation on FPGAs is performed.

The work by Kasap~et~al.~\cite{Kasap2012} is another attempt to implement a production grade MD accelerator using FPGAs which is not from the Herbordt group. They target the Maxwell platform, an FPGA based multi-node system that has similarities with Microsoft's Catapult \cite{putnam2014reconfigurable,caulfield2016cloud} (albeit using more dated technology, i.e. Virtex 4 FPGAs). The overall system architecture template is very similar to the one used in this report. However, they only accelerate non-bonded pair interactions on the FPGAs and do not use inter-FPGA communication. Although they are able to accelerate the interactions, the overall system is not competitive due to a very limited bandwidth between host processor and the SD RAM on the FPGA card (PCI-X bus, with only around \SI{500}{\mega\byte\per\second} for two FPGA cards). Further, they transfer many parameters that could be shared among several particles on a per atom basis onto the the accelerator which leads to significant I/O overhead. Virials and potential energies are calculated in each iteration, which is typically not required. Due to these factors, their FPGA solution is actually slower than the software baseline, since about 96\% of the time is spent in communication. Apart from the apparent improvements in data communication volume (redundant parameters), the situation in connectivity is quite different today, where PCIe and NVLink provide much higher bandwidth between host and device. Further, modern FPGAs provide enough fast on-chip RAM resources such that no external SRAM/DRAM has to be used (this is another performance bottleneck of the system by Kasap~et~al.).

In general, we can observe that FPGA designs have to leverage fixed-point arithemtic and operator fusing wherever possible (e.g. in the interpolators and for particle coordinates). Floating-point arithmetic should only be used where absolutely needed (e.g., final force values) as these cause high pipeline latencies and require a lot of LUT resources. According to literature \cite{Chiu2011efficient}, the limiting resource types on modern FPGAs are likely going to be the  LUTs and registers in this context, and the amount of DSP slices and memory blocks are of secondary concern. Further, fixed-point arithmetic enables to reach higher clock rates and maps better to DSP resources than FP. Note that the Anton chips almost exclusively employ fixed-point datapaths, too, since the use of fixed-point not only improves performance, but also has a positive impact on testability and reproducibility (out-of-order accumulations remain bit-true, for instance).

We base the FPGA estimates for the PP interaction and PME modules in this report on the resource figures provided by Herbordt's group, as these appear to be the best references that can be found in literature.
\subsection{High-Level Synthesis}
High-level synthesis (HLS) and related methods for FPGAs are often advertised as great productivity enhancers that significantly reduce the complexity of design-entry compared to traditional HDL flows. From experience, we can say that this is not always the case, as the benefit of HLS is quite design dependent. However, the results presented by \cite{sanaullahopencl,yang2017opencl,Cong2016,waidyasooriya2017fpga} suggest that HLS or OpenCL based flows can indeed be used for certain parts of the PP and PME units.

For example, Saunullah~et~al.~\cite{sanaullahopencl} compare an IP-based design flow with FFT IP's from Altera to an OpenCL kernel, and report quite large improvements in terms of overall resources (10$\times$ fewer ALMs, 25$\times$ less on-chip memory). While this needs to be interpreted with care (the paper does not reveal many implementation details), it seems to be reasonable that such an approach yields good results in this setting. The FFT is quite a regular algorithm which can be well described with OpenCL, and the Altera FFT IP's incur some overhead due to internal buffering, interfaces, etc. which are not needed.

On the other hand, Saunullah~et~al.~\cite{yang2017opencl} attempt to use OpenCL to design the same PP interaction pipeline that they developed with HDL in an earlier papers \cite{Chiu2009efficient,Chiu2010towards,Chiu2011efficient}. Their conclusion is that for such a highly tuned datapath OpenCL does not provide competitive results ("...the OpenCL versions are dramatically less efficient, with the Verilog design fitting from 3.5$\times$ to 7$\times$ more logic."). The work by Cong~et~al.~\cite{Cong2016} is a similar study that investigates a Xilinx HLS flow for the same PP interaction module. Their experiments reveal that part of the inefficiency of HLS for this particular module is the fact that HLS tools currently have difficulties producing efficient results for datapaths with dynamic data flow behavior where conditional execution exists within a processing element. Their solution is to describe certain parts critical for dynamic data flow in HDL, and generate the remaining subblocks of the datapath using HLS.

So based on our own experience and the above literature, we can say that
OpenCL or HLS can increase productivity, and yield good results for algorithms that can be described well and that have regular compute patterns (e.g., systolic dataflow, no feedback loops, no dynamic control behavior). But for designs that are difficult to describe and tune with OpenCL such as parts of the force-pipeline a highly tuned HDL implementation turns out to be more efficient. An interesting design approach is to use a mix between both methodologies, leveraging HLS (and its automated verification functionality) for small datapath subblocks that are then connected and orchestrated using an HDL wrapper.
\begin{table}[!t]
 \caption{Test nodes used for benchmarking. The P[2,3].[2,8,16]xlarge nodes are compute instances available on AWS.}
 \label{tab:machinesTable}
 \renewcommand{\arraystretch}{1.0}
 \centering
 \small
 \begin{threeparttable}
 \begin{tabularx}{0.48\textwidth}{@{}p{2cm}|c|c|l@{}}
 \toprule
 \textbf{Node Name} & \textbf{CPU (GHz)} & \textbf{vCPUs}$^\dag$ &  \textbf{GPU Config}\\
 \midrule
 Sassauna           & E5-2640v4 (2.4) & 40 & 4$\times$GTX1080Ti  \\
 P2.2xlarge         & E5-2686v4 (2.3) & 8  & 1$\times$K80$^\ddag$\\
 P2.8xlarge         & E5-2686v4 (2.3) & 32 & 4$\times$K80$^\ddag$\\
 P2.16xlarge        & E5-2686v4 (2.3) & 64 & 8$\times$K80$^\ddag$\\
 P3.2xlarge         & E5-2686v4 (2.3) & 8  & 1$\times$V100 SMX2  \\
 \bottomrule
 \end{tabularx}
 \begin{tablenotes}
 \item[$^\dag$] The number of vCPUs (virtual CPUs) corresponds to the amount of concurrent threads.
 \item[$^\ddag$] The K80 is a dual GPU card, so the effective number of GPUs is double the listed number.
 \end{tablenotes}
 \end{threeparttable}
\end{table} 
\subsection{Vendor and Market Overview}
Several well known high-performance software packages come from the academic sector and some of them like  GROMACS \cite{Hess2008a,Kutzner2015,Abraham2015a} and LAMMPS \cite{plimpton1995} are released freely under open-source licences. Others like such as AMBER \cite{amber2018}, NAMD \cite{phillips2005scalable} and CHARMM \cite{brooks2009charmm} provide reduced or free academic licenses and require full licensing for commercial purposes.

Besides these codes originating from academia, there are a couple of companies that develop and sell MD simulation software such as Biovia from Dassault Systemes, AceMD from Acellera Ltd \cite{Harvey2009}, Yasara \cite{Krieger2015}, and Desmond from Shaw Research LLC \cite{Bowers2006} to name a few (see this market report for more info \cite{grandView2017}). Some of the main differentiators of these commercial software packages comprise:
\begin{itemize}
\item Workflow integration and collaborative tools (e.g., Dassault Systems),
\item Better visualization and GUI integration (e.g., Dassault Systems, Yasara),
\item Performance tweaks for workstation systems (e.g., Yasara, Acellera),
\item Extreme scalability on clusters (e.g., Shaw Research),
\item Costumer support and consulting options (all of them).
\end{itemize}
In terms of infrastructure, some companies sell \emph{application certified} workstations and server blades, such as ExxactCorp\footnote{\url{https://www.exxactcorp.com/GROMACS-Certified-GPU-Systems}} (GROMACS and AMBER workstations) and Acellera (AceMD MetroCubo)\footnote{\url{https://www.acellera.com/}}. Another interesting trend in this field seem to be cloud services \cite{grandView2017,insightPharma2010}. E.g., Acellera AceMD has built-in support for Amazon Web Services (AWS), that allows users to conveniently outsource simulation runs with a single command. The advantage of such cloud solutions are scalability and low up-front costs, which can be attractive for small labs and/or labs that have large variations in workload. Otherwise on-prem solutions can be more cost-effective.

The overall market however can be considered a niche market, since there are only a few 1000 to 10'000 users worldwide that use such specialized software packages. According to Goldbeck~et~al.~\cite{goldbeck2017}, the overall spending on scientific software is in the order of 0.1\% of the total sector R\&D spending, which would amount to roughly 100M\$ in pharma/biotech and 50M\$ in the chemicals/materials industry in Europe. Now this is an estimate for the total spending, so for a specific software package the market size will only be a fraction of that. So assuming a share in the order of 1\% we can estimate that the market for a specific software package is likely in the order of several 100k\$ to a few M\$.

In this report, we use AMBER and GROMACS as benchmark baselines, since these provide very competitive runtimes on single-node systems and are widely used in the community. In addition to on-prem solutions, we also consider cloud infrastructure, since FPGAs have recently become available as a service (FaaS), in the form of the AWS F1 instance\footnote{\url{https://aws.amazon.com/ec2/instance-types/f1/}}. Edico Genomics\footnote{\url{http://edicogenome.com/}} is an example for a company that successfully uses F1 instances to accelerate Genome sequencing.
\section{Software Benchmarks}
\label{sec:bench}
In this section we assess the performance of two widely used GPU-accelerated MD packages using three different benchmarks and recent hardware. This is done to get a solid baseline for the comparisons with the FPGA performance estimates in \secref{sec:perfEst}.
\subsection{MD Packages and Hardware Configuration}
We use GROMACS 2018.1 and a licensed version of AMBER 16 (with AmberTools 17 and all patches that where available by the end of June 2018) in this report, since these are widely used and are among the fastest software packages for single node systems. Both packages have been compiled from scratch on Linux 14.06 LTS with GCC 5.5.0. GROMACS has been compiled in mixed-precision mode for single node systems using the built-in thread-MPI support and with FFTW 3.3.5 and CUDA 9.2. AMBER has been built in mixed-precision as well with MPI support, FFTW 3.3 and CUDA 9.0.

We used one on-site machine and several different Amazon Webservices (AWS) configurations to run the benchmarks. The machine configurations are listed in \tabref{tab:machinesTable}. All machines have almost identical processor models (only the core count and maximum frequency differs slightly), and all machines ran Linux 14.06 LTS. The configuration of the on-site machine \emph{Sassauna} is in line with the GROMACS benchmarking paper by Kutzner~et~al.~\cite{Kutzner2015best}, where it has been found that a dual Xeon machine with around 10 cores per socket and 2-4 customer grade GPUs provides the best cost-efficiency in terms of \SI{}{\nano\second}/\$.
\begin{figure*}[!ht]
\centering
\subfigure[]{
\includegraphics[width=0.425\textwidth]{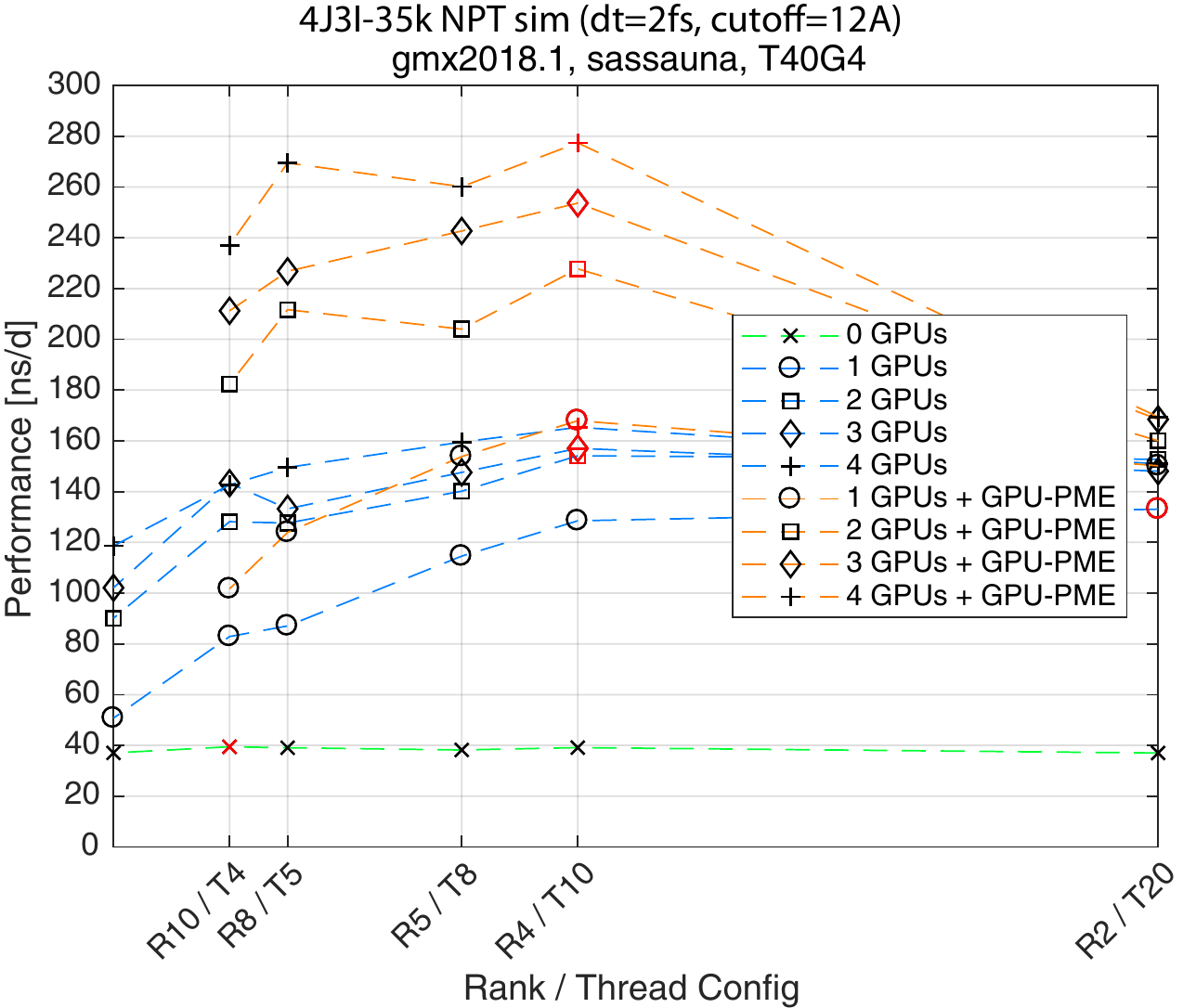}
}
\subfigure[]{
\includegraphics[width=0.425\textwidth]{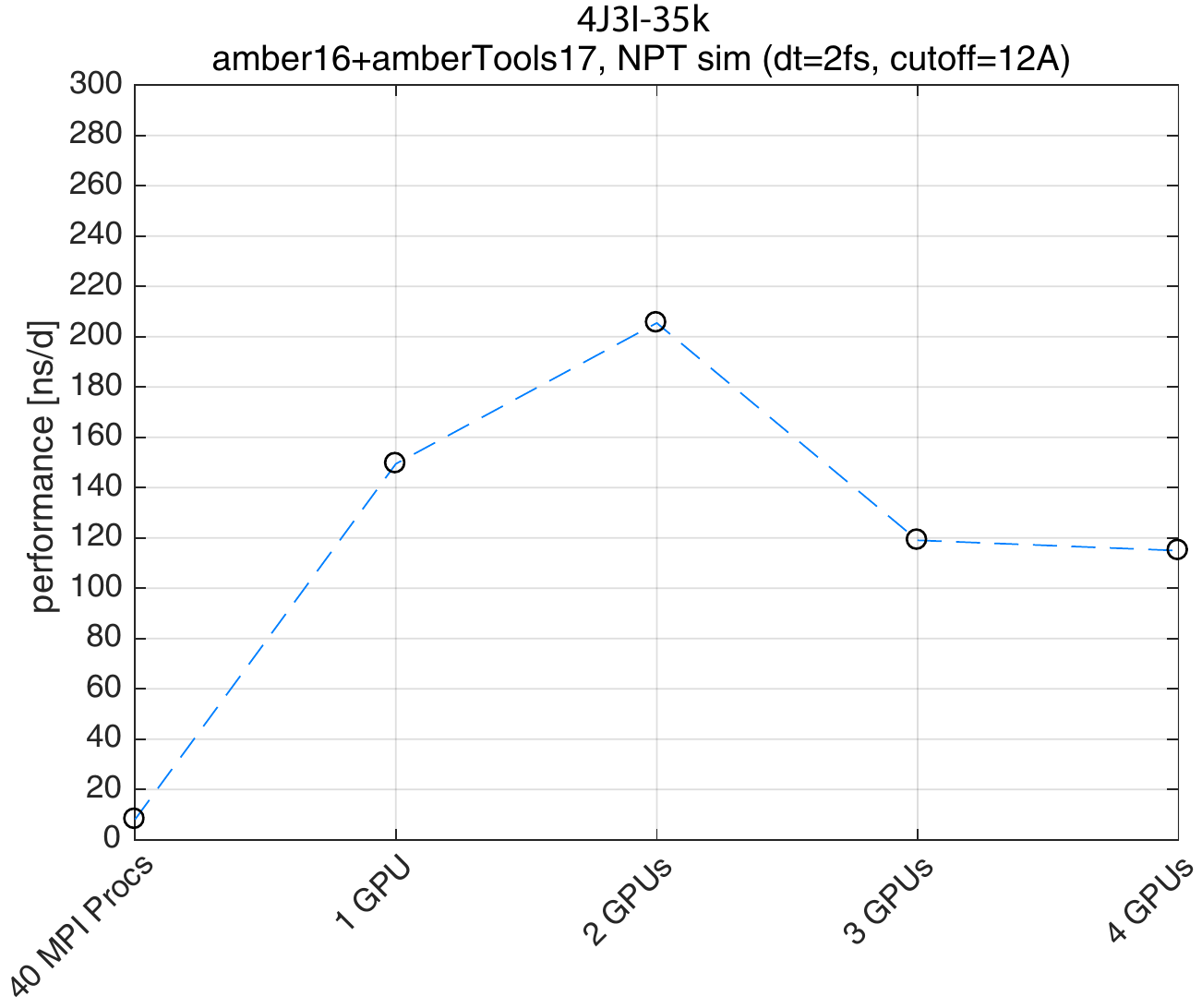}
}
\\
\subfigure[]{
\includegraphics[width=0.425\textwidth]{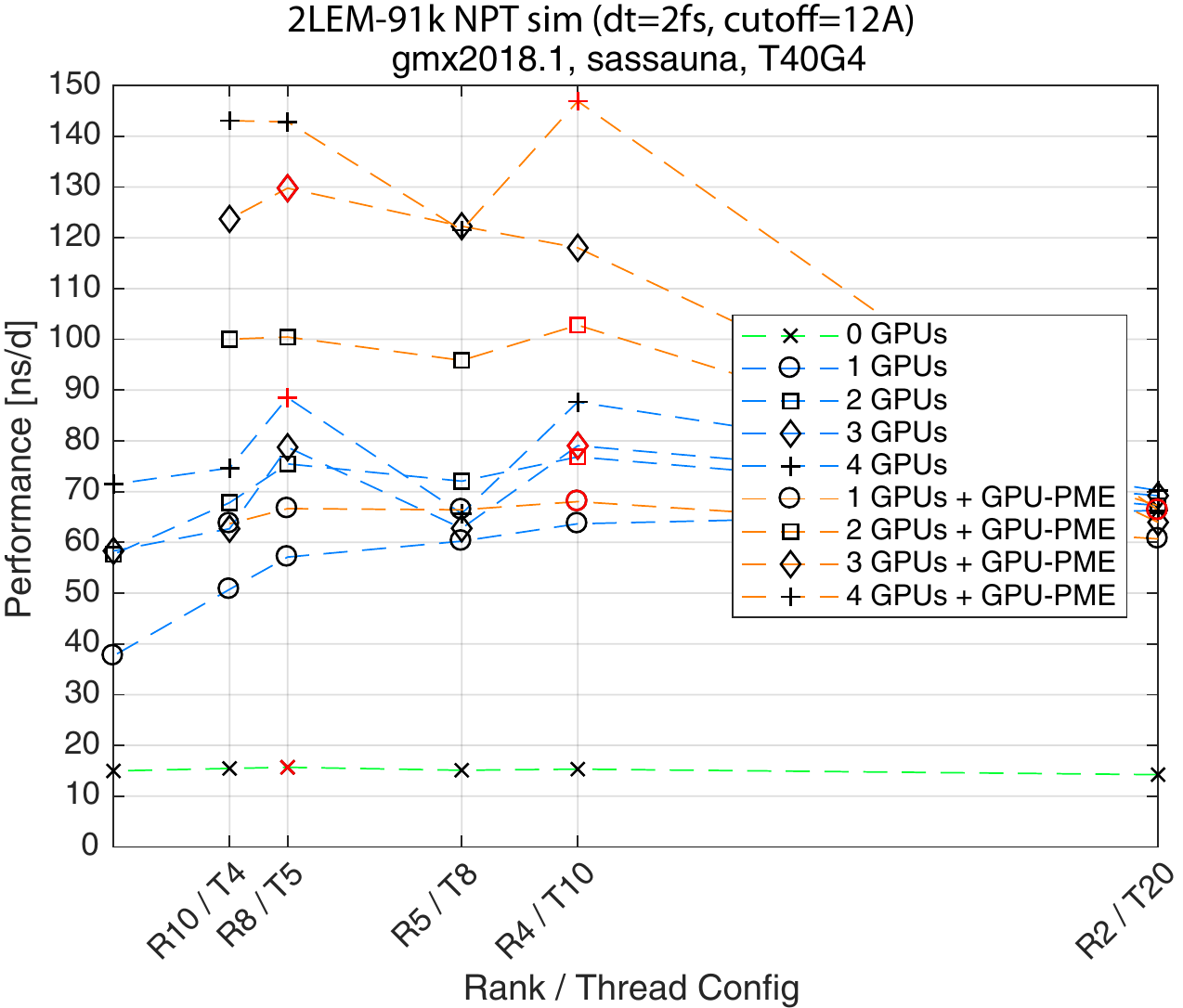}
}
\subfigure[]{
\includegraphics[width=0.425\textwidth]{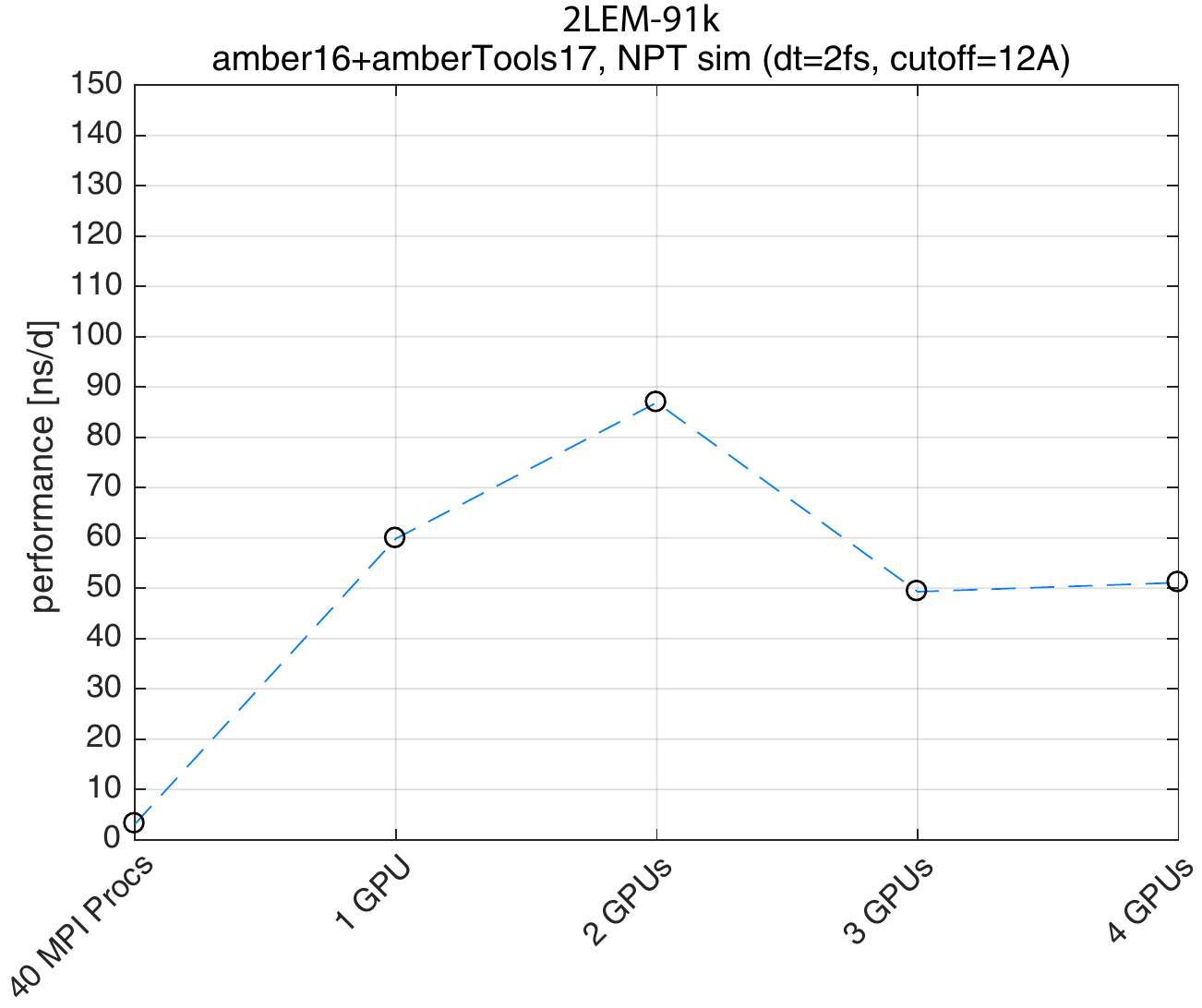}
}
\\
\subfigure[]{
\includegraphics[width=0.425\textwidth]{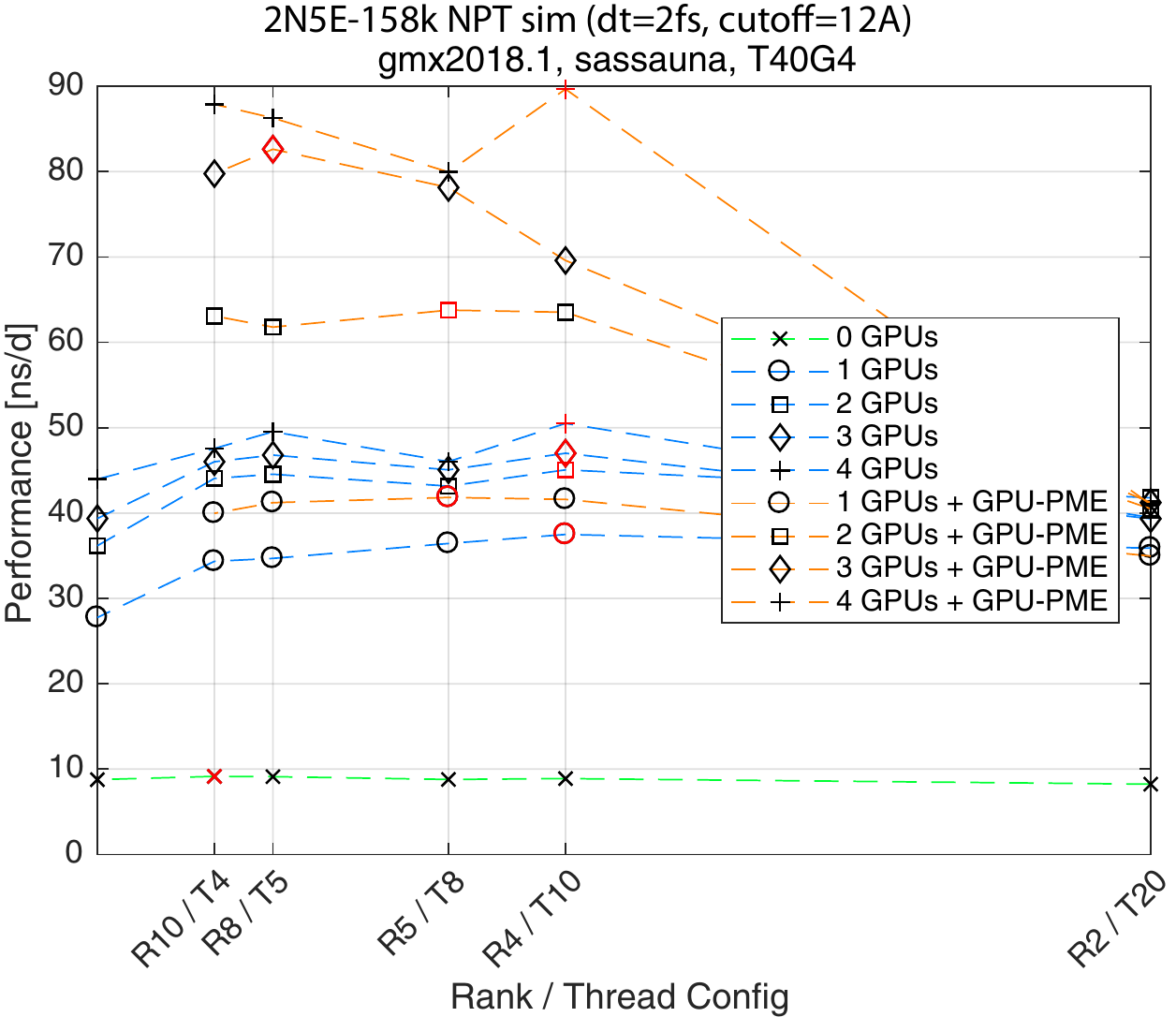}
}
\subfigure[]{
\includegraphics[width=0.425\textwidth]{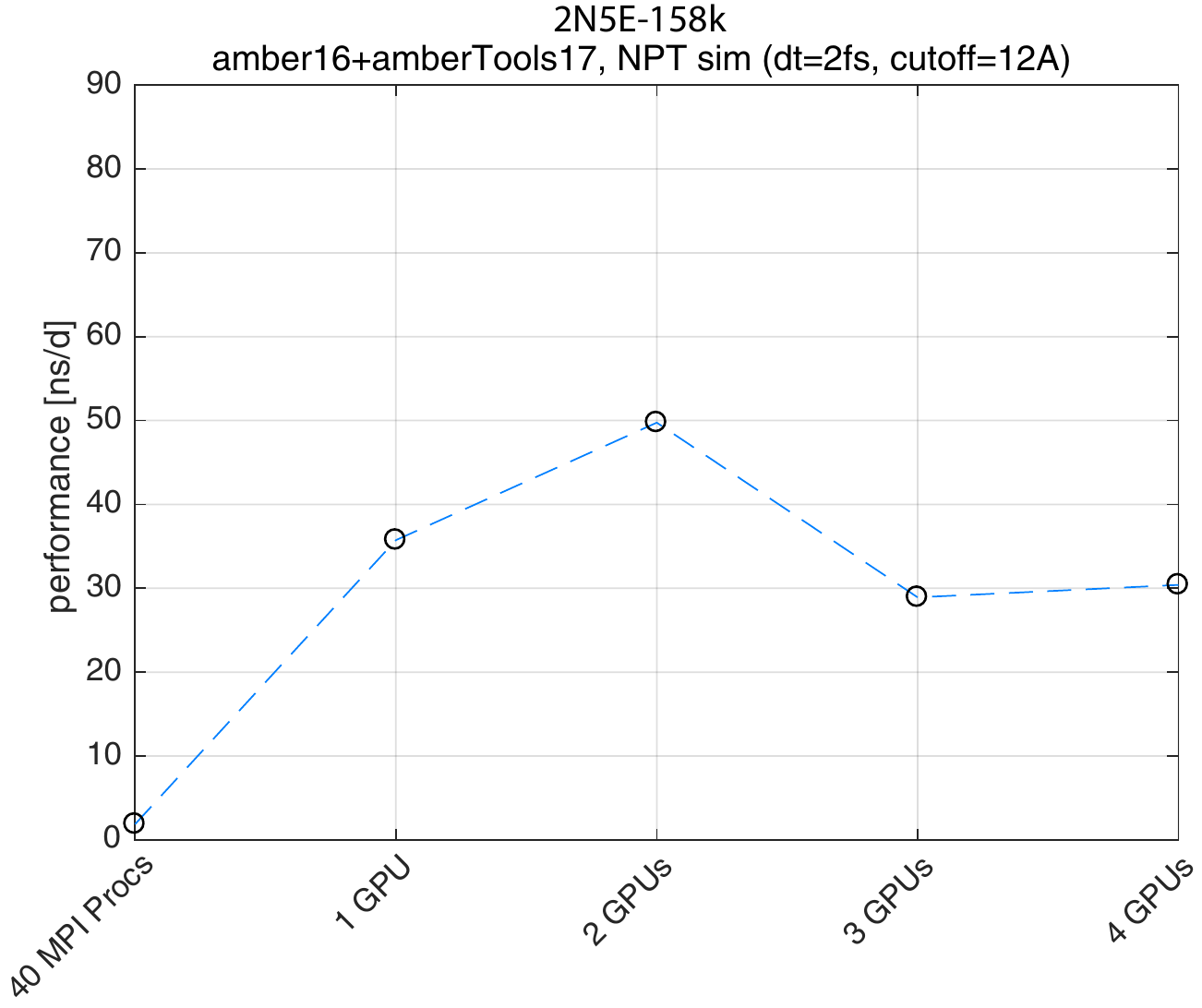}
}
\\
\caption{Performance of AMBER and GROMACS with different \#thread vs. \#GPU configurations on the Sassauna machine.}
\label{fig:gmxAmberPerf}
\end{figure*}
\begin{figure*}[!t]
\centering
\subfigure[]{
\includegraphics[width=0.3\textwidth]{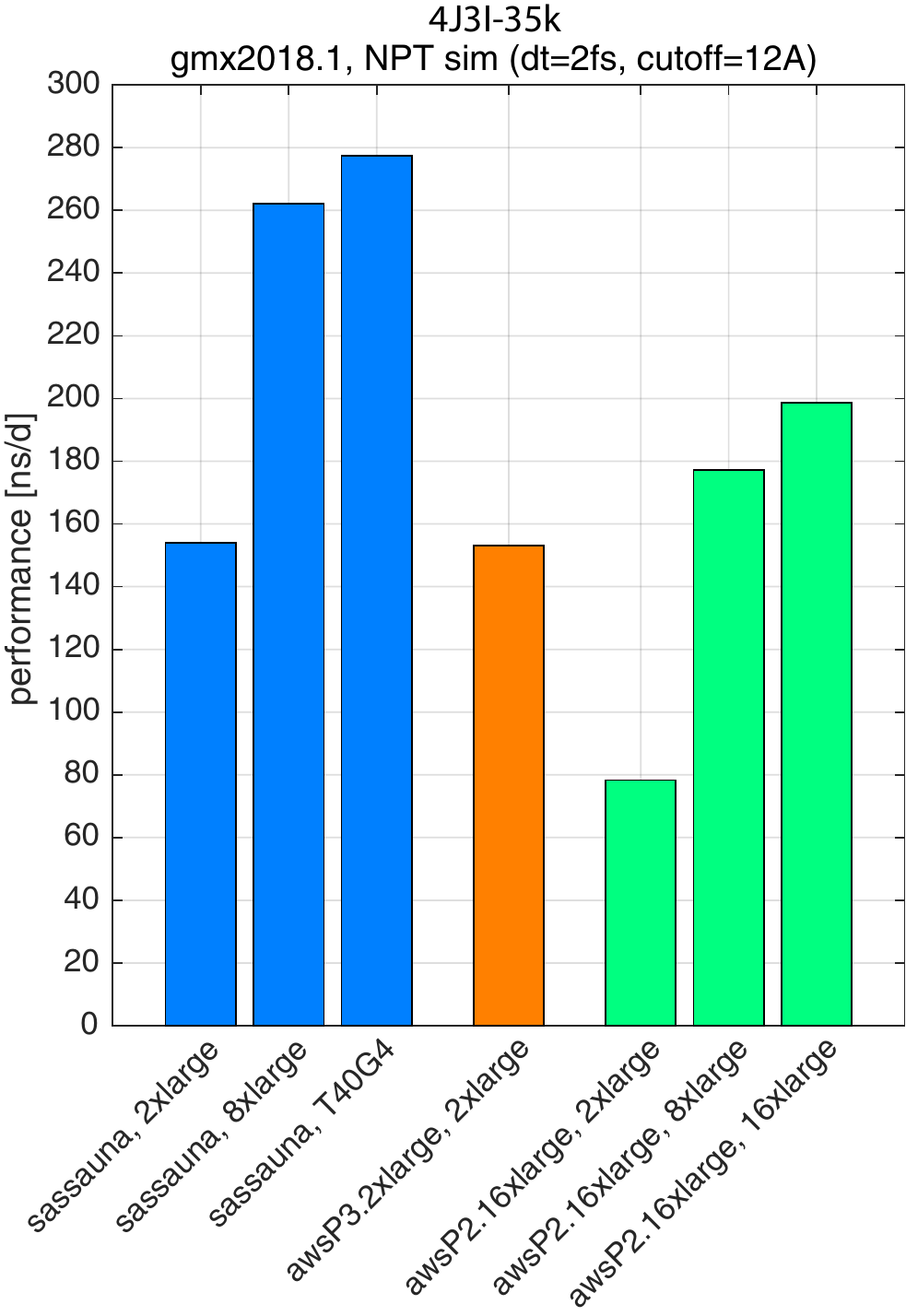}
}
\subfigure[]{
\includegraphics[width=0.3\textwidth]{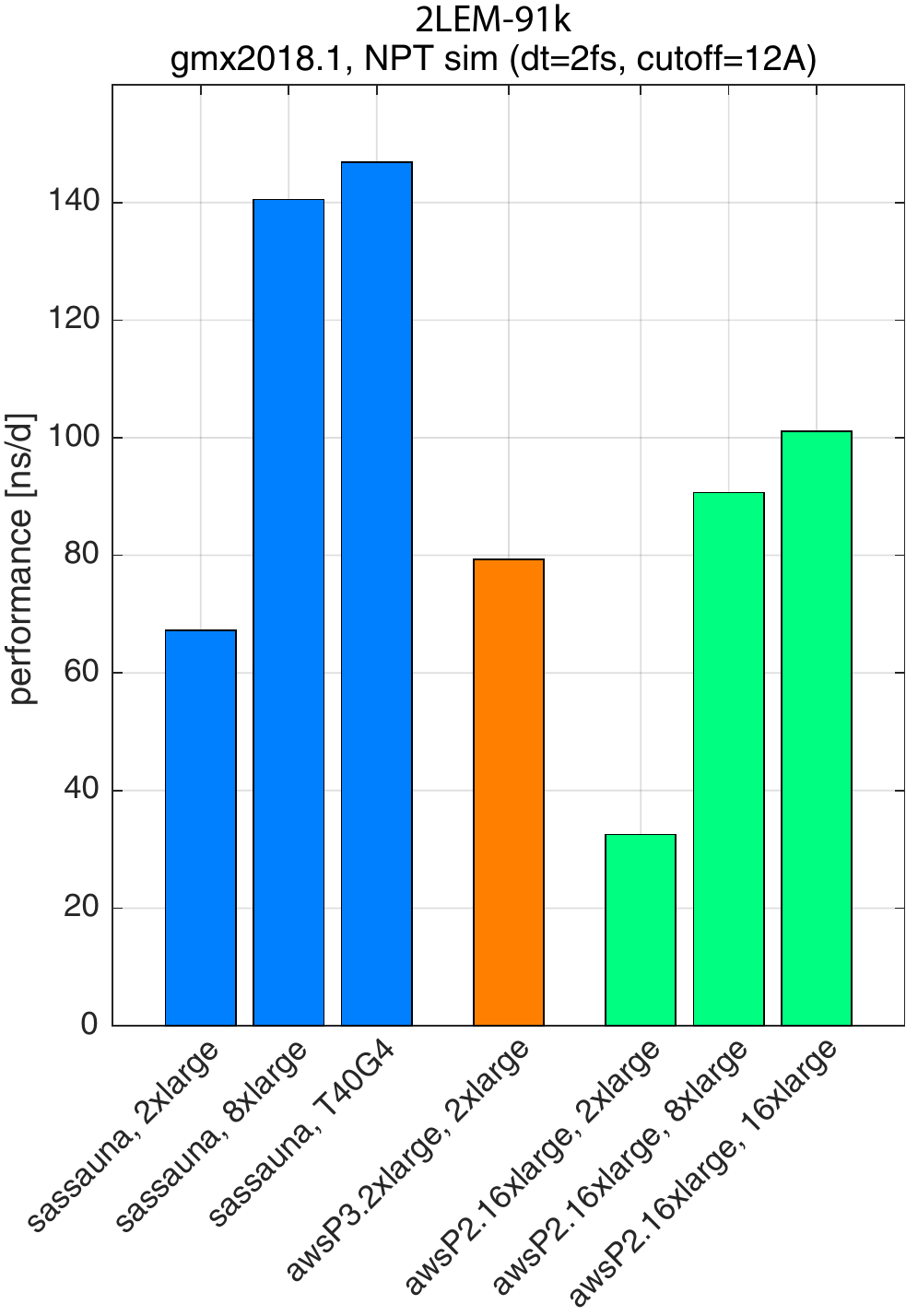}
}
\subfigure[]{
\includegraphics[width=0.27\textwidth]{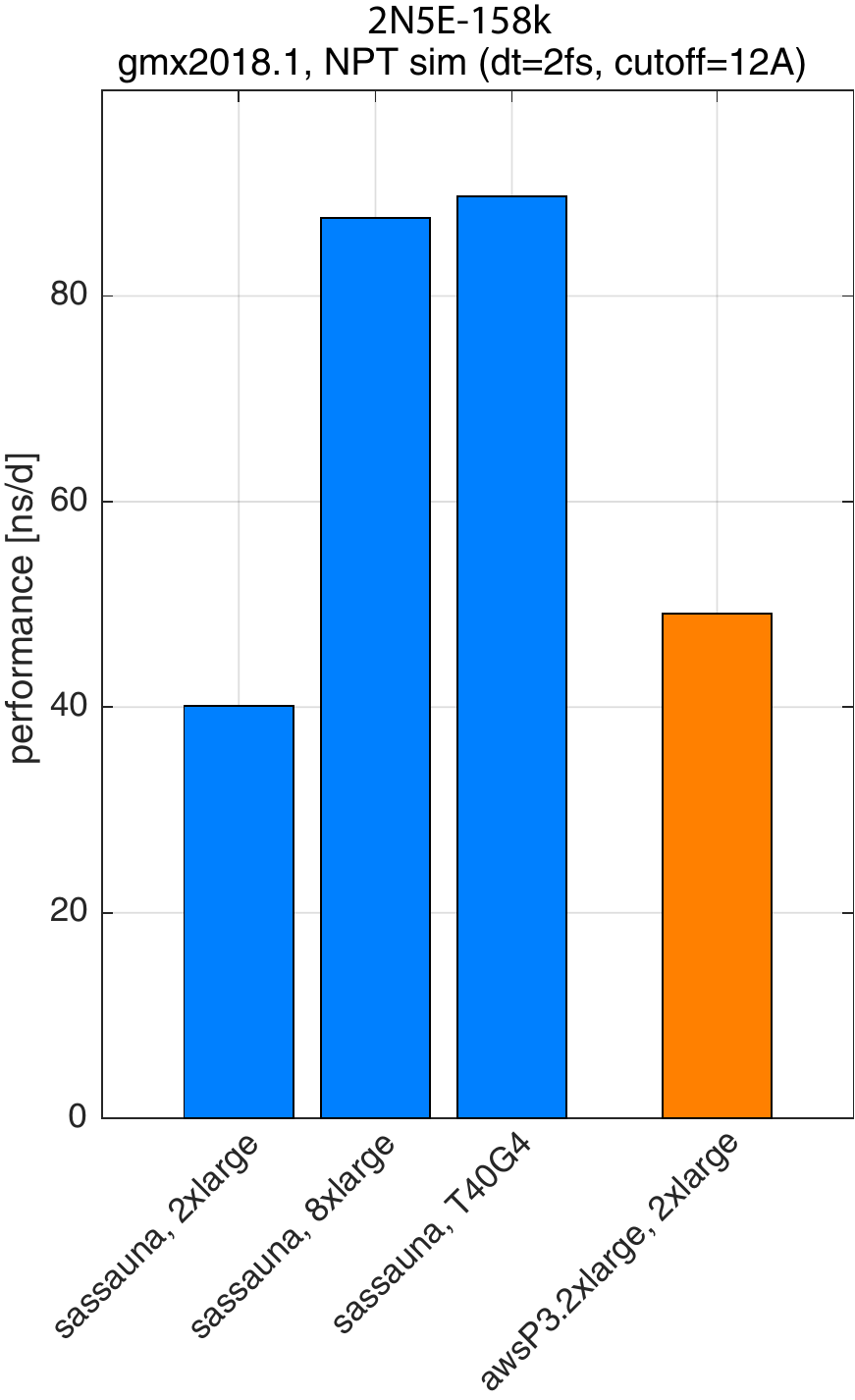}
}
\\
\caption{Comparison of the performance of GROMACS on different AWS instances and the corresponding configurations on the Sassauna machine. The largest benchmark has not been run on AWS due to long benchmark runtimes.}
\label{fig:awsComparison}
\end{figure*}
\subsection{Benchmark Systems}
\begin{table}[!t]
 \caption{MD systems used for benchmarking.}
 \label{tab:proteinTable}
 \renewcommand{\arraystretch}{1.0}
 \centering
 \small
 \begin{threeparttable}
 \begin{tabularx}{0.48\textwidth}{@{}p{3cm}|c|c|c@{}}
 \toprule
                            & \includegraphics[width=1.5cm]{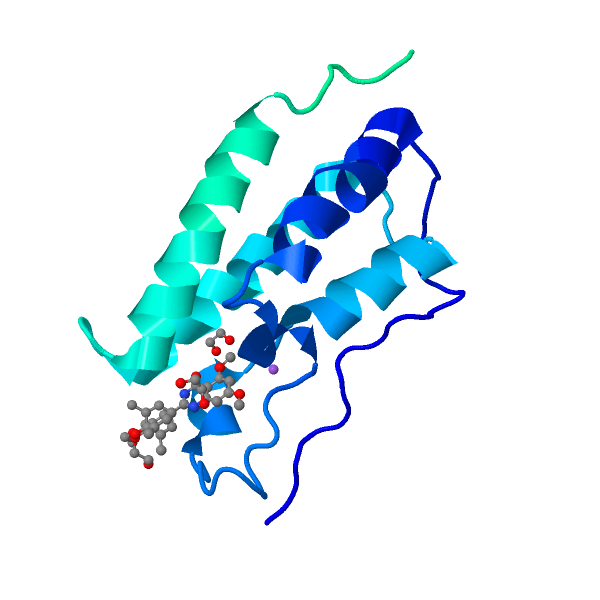} & \includegraphics[width=1.5cm]{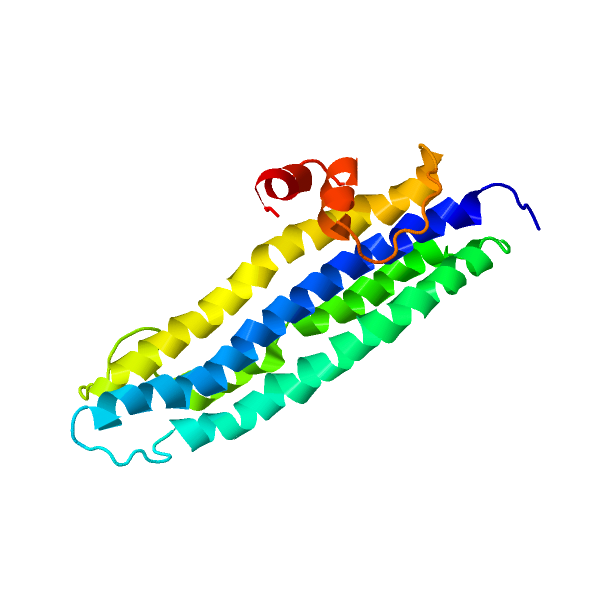} & \includegraphics[width=1.5cm]{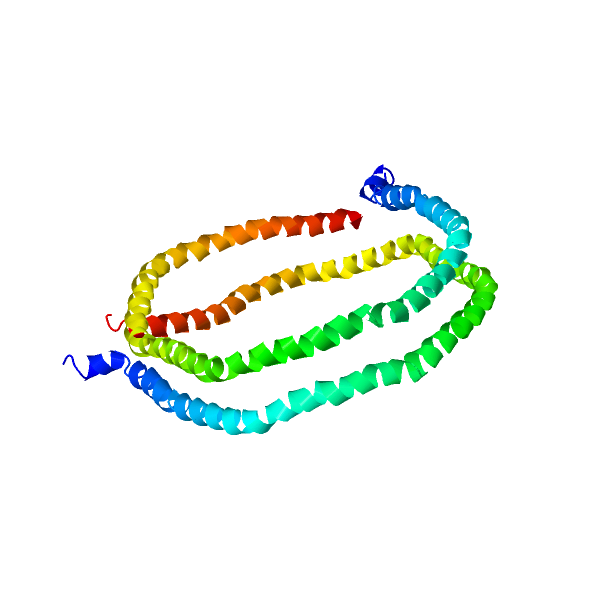} \\
  \midrule
  \textbf{Protein ID}       & \textbf{4J3I} & \textbf{2LEM} & \textbf{2N5E} \\
  \midrule
  \# Particles              & 35'022 & 91'030 & 157'853 \\
  System Size [\AA]         & $\texttildelow$72$^3$ & $\texttildelow$99$^3$ & $\texttildelow$119$^3$ \\
  Time Step [fs]            & 2      & 2      & 2       \\
  Cutoff Radii [\AA]$^\dag$ & 12     & 12     & 12      \\
  PME Grid$^\dag$           & 60$^3$ & 84$^3$ & 100$^3$ \\
  \midrule
  PME Tuning Steps$^g$      & 6k     & 6k     & 6k      \\
  Benchmark Steps$^g$       & 9k/14k$^\ddag$ & 9k/14k$^\ddag$ & 9k/14k$^\ddag$  \\
  Benchmark Steps$^a$       & 10k    & 10k    & 10k     \\
  \bottomrule
 \end{tabularx}
 \begin{tablenotes}
 \item[$\dag$] These are the initial values before load balancing, see also \cite{Kutzner2015best}.
 \item[$\ddag$] CPU/GPU runs.
 \item[$g$] GMX runs.
 \item[$a$] AMBER runs.
 \end{tablenotes}
 \end{threeparttable}
\end{table} 
Existing benchmarks are often MD software specific (in terms of input files), and hence we chose to recreate the input files from scratch to make sure we are simulating the same system. To accomplish that we used the CHARMM-GUI\footnote{Input generator on \url{http://www.charmm-gui.org/}} online tool \cite{Sunhwan2008,Lee2016}. The benchmarks are listed in \tabref{tab:proteinTable}, and consist of three different proteins that have been solvated in a box of water. The amount of particles range from 35k to 157k, which represents typical problem sizes in biochemistry today. The two smaller systems are similar to the DHFR and Apoa1 benchmarks often used in literature in context with the AMBER and NAMD software packages.
\subsection{Benchmarking Results}
We use the standard protocols provided by CHARMM-GUI to equilibrate the systems. The equilibrated systems are then simulated in the NPT ensemble for the amount of steps specified in \tabref{tab:proteinTable}. In GROMACS, a Nose-Hoover thermostat is used in combination with Parinello-Rahman pressure control and in Amber, a Langevin Thermostat is used together with Montecarlo Pressure control (these were the default settings from CHARMM-GUI). The force-fields employed is CHARMM36m, which uses a TIPS3P water model (with Coulomb and LJ interactions on the O and H atoms). Energies are calculated every 100 and stored every 1k steps in both cases. For each software package, we run different parallelization options to find the optimal one.
\subsubsection{GROMACS Benchmarks on Sassauna}
Similar as described in \cite{Kutzner2015best}, we sweeped different MPI rank/openmp thread combinations, in combination with different numbers of GPUs. The results are shown in \figref{fig:gmxAmberPerf}a,c,e. For standard GPU accelerated runs (blue lines) where the particle-particle (PP) interactions are mapped to the GPUs and the PME is handled on the CPUs\footnote{This is done without separate PME ranks in this case.}, we observe speedups in the order of $4\times$ to 6$\times$ with respect to the CPU-only version (green line). Note that in GROMACS 2018 update, a new feature became available that allows to map the PME evaluation to one of the GPUs, which leads to significant runtime improvement (see orange lines) on single-node systems as the costly 3D FFT communication phases are now contained within one device instead of being spread accross many CPU cores. This leads to a performance improvement of another $2\times$.
\subsubsection{AMBER Benchmarks on Sassauna}
As can be seen in \figref{fig:gmxAmberPerf}b,d,f, the CPU-only implementation is quite slow when compared to GROMACS. However, the GPU accelerated configuration with 2 GPUs provides similar performance as GROMACS with 1-2 GPUs using the new GPU PME feature. For more GPUs, the performance does not scale on this system. Note that as opposed to GROMACS, AMBER can run the whole simulation loop on the GPUs without involving the CPU cores. In multi-GPU runs, several GPUs communicate via the PCIe bus. This is only working well for up to 2 cards in this machine, since each pair is connected to one XEON socket, and connections bettween these pairs have to go over QPI. Compute nodes with NVLINK or optimized PCIe infrastructure will likely perform better on such runs\footnote{See \url{http://ambermd.org/gpus/benchmarks.htm}, \url{http://ambermd.org/gpus}}. The advantage of AMBER is not necessarily the speed of a single MD run (GROMACS is faster in that case), but the fact that the no expensive multi-core CPUs have to be used in order to get decent performance as in GROMACS. This allows to run many simulations in parallel in a very cost-effective manner.

Note that we used the licensed version of AMBER16 for these benchmarks with all updates as of June 2018. There also exists AMBER18 that has been released recently, and that likely includes further optimizations for the Pascal and Volta generation GPUs.
\subsubsection{GROMACS Benchmarks on AWS}
As mentioned in the introduction, cloud infrastructure/software services (IaaS/SaaS) are interesting alternatives to on-site solutions. Hence, we also benchmark a few AWS instances in order to be able to include them in the performance cost comparison later on in \secref{sec:perfEst}.

At the time of writing, the demand for P3 instances with V100 SMX2 cards was extremely high, and hence we only got access to one 2xlarge instance with one GPU. As such, the performance improvements for multi-GPU runs with V100's has to be extrapolated from these measurements. As we can observe in \figref{fig:awsComparison} the improvement is in the order of 20\% with respect to the GTX1080Ti for the larger benchmarks. This improvement seems to be reasonable, since the raw increase in FLOP/s is around 35\%. We can also see that the P2 instances with K80 GPUs are significantly slower than the P3 instance or Sassauna.
\subsubsection{GROMACS Breakdown of Simulation Loop}
\begin{table}[!h]
 \caption{Wall-time accounting as reported by GROMACS for different configurations (in percent).}
 \label{tab:breakdownTable}
 \renewcommand{\arraystretch}{1.0}
 \centering
 \small
 \begin{threeparttable}
 \begin{tabularx}{0.48\textwidth}{@{}p{3cm}|c|c|c@{}}
 \toprule
 \multicolumn{4}{c}{\textbf{Single Threaded}}\\
 \midrule
 \textbf{Compute Steps}     & \textbf{4J3I-35k}   & \textbf{2LEM-91k}   & \textbf{2N5E-158k} \\
 Force                      &  87.7               &   86.2              &  86.3         \\
 PME mesh                   &  8.4                &    9.7              &   9.9         \\
 NB X/F buffer ops.         &  0.2                &    0.4              &   0.4         \\
 Write traj.                &  2.9                &    3.0              &   2.6         \\
 Update                     &  0.2                &    0.3              &   0.3         \\
 Constraints                &  0.4                &    0.4              &   0.4         \\
 Rest                       &  0.1                &    0.1              &   0.1         \\
 \midrule
 %\multicolumn{4}{c}{}\\
 %\midrule
 \multicolumn{4}{c}{\textbf{Multi Threaded (40T), PP Accelerated (4 GPUs)}}\\
 \midrule
 \textbf{Compute Steps}     & \textbf{4J3I-35k}   & \textbf{2LEM-91k}   & \textbf{2N5E-158k} \\
 Domain decomp.             &  1.7                &  2.1                &  2.1 \\
 DD comm. load              &  0.0                &  0.0                &  0.0 \\
 DD comm. bounds            &  0.0                &  0.1                &  0.0 \\
 Neighbor search            &  1.2                &  1.4                &  1.4 \\
 Launch GPU ops.            & 14.2                &  9.5                &  5.9 \\
 Comm. coord.               &  9.4                & 12.2                &  8.9 \\
 Force                      &  3.6                &  2.8                &  2.9 \\
 Wait + Comm. F             & 10.0                & 10.0                &  9.1 \\
 PME mesh                   & 44.8                & 46.9                & 52.6 \\
 Wait GPU NB nonloc.        &  0.8                &  0.3                &  0.6 \\
 Wait GPU NB local          &  0.4                &  0.3                &  0.3 \\
 NB X/F buffer ops.         &  2.8                &  3.4                &  4.3 \\
 Write traj.                &  1.1                &  0.5                &  0.9 \\
 Update                     &  0.9                &  1.0                &  2.5 \\
 Constraints                &  7.0                &  6.2                &  6.7 \\
 Comm. energies             &  1.1                &  2.4                &  0.8 \\
 Rest                       &  1.0                &  0.7                &  0.9 \\
 \midrule
 %\multicolumn{4}{c}{}\\
 %\midrule
 \multicolumn{4}{c}{\textbf{Multi Threaded (40T), PP+PME Accelerated (4 GPUs)}}\\
 \midrule
 \textbf{Compute Steps}     & \textbf{4J3I-35k}   & \textbf{2LEM-91k}   & \textbf{2N5E-158k} \\
 Domain decomp.             &  3.2                &  3.9                & 3.7   \\
 DD comm. load              &  0.0                &  0.0                & 0.0   \\
 Send X to PME              &  4.7                &  5.9                & 6.2   \\
 Neighbor search            &  1.9                &  2.1                & 2.1   \\
 Launch GPU ops.            &  14.4               &  7.9                & 5.0   \\
 Comm. coord.               &  14.0               &  14.7               & 14.1  \\
 Force                      &  6.4                &  5.3                & 4.6   \\
 Wait + Comm. F             &  15.2               &  13.3               & 11.0  \\
 Wait + Recv. PME F         &  3.7                &  4.9                & 8.8   \\
 Wait PME GPU gather        &  8.7                &  13.1               & 14.6  \\
 Wait GPU NB nonloc.        &  6.5                &  7.6                & 7.3   \\
 Wait GPU NB local          &  0.6                &  0.3                & 2.2   \\
 NB X/F buffer ops.         &  4.3                &  4.6                & 4.3   \\
 Write traj.                &  2.2                &  1.9                & 2.0   \\
 Update                     &  2.2                &  3.0                & 3.4   \\
 Constraints                &  11.8               &  11.3               & 10.6  \\
 Comm. energies             &  0.1                &  0.1                & 0.1   \\
 \bottomrule
 \end{tabularx}
 \begin{tablenotes}
 %\item[$\dag$]
 %\item[$\ddag$]
 \end{tablenotes}
 \end{threeparttable}
\end{table} 
The walltime breakdown for the three different benchmarks is shown in \tabref{tab:breakdownTable} in \% for three different cases: single threaded (CPU-only), multi-threaded with PME on CPUs, and multi-threaded with both PP interactions and PME on GPUs.

The walltime breakdown for the single-threaded case is shows the well-known picture: the compute time is mainly dominated by non-bonded force evaluations, which accounts for more than 96\%, including PME. The force time also includes bonded forces in this case, but the fraction is insignificant compared to the non-bonded part.

In the multi-threaded case with PP interactions on the GPU, the picture already changes quite a bit. The runtime of the PP interaction kernels is not visible since they are executed in parallel to the CPU code, for which the walltime accounting is performed. The breakdown is a bit more difficult to read, but we see that PME starts to become the dominant factor (more than 50\%)..

In the third case, we note that other parts besides the PP interactions and PME are starting to become significant as well. In particular the operations that cannot be overlapped with the accelerated PP and PME calculations (such as domain decomposition, reduction operations, trajectory sampling, integration, global energy communication) start to amount for a significant percentage of overall walltime (around 25\% in this benchmarks). As we shall see in \secref{sec:perfEst} on hardware performance estimates, this inherently limits the maximum acceleration that can be achieved by using a co-processor solution. This issue is not exclusive to GROMACS, and has also been noted, e.g., in \cite{Hardy2018} with NAMD.

It is worth noting that the constraints step is required to keep high-frequency oscillations of bonds involving light atoms (H) under control. Otherwise, significantly smaller timesteps than the currently employed \SI{2}{\femto\second} have to be employed in order to ensure correct integration and prevent the simulation from \emph{blowing up}. These constraints are implemented using a combination of efficient parallel algorithms in GROMACS (P-LINCS for general bonds \cite{Hess1997,Hess2008}, analytic SETTLE \cite{Miyamoto1992} for water molecules).
\section{Performance Estimations and Comparisons}
\label{sec:perfEst}
In this section, we estimate the performance of FPGA accelerated systems, compare them with GPU-based solutions in terms of performance/cost and discuss the technical and commercial implications.
\subsection{Considered System Architectures}
\begin{figure}[t]
\centering
\includegraphics[width=0.48\textwidth]{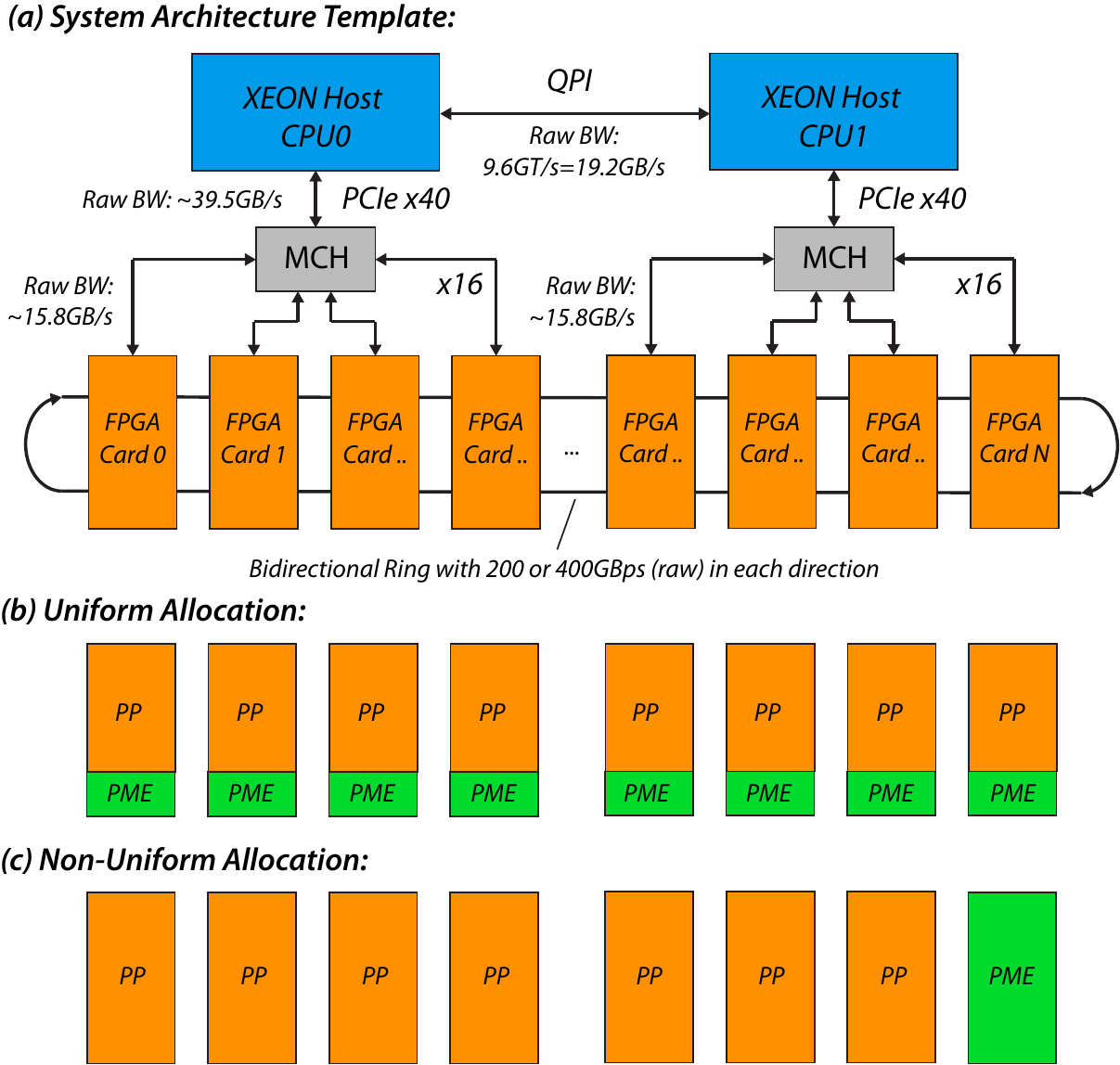}
\caption{Architectural template used for the performance estimations and different PP/PME unit allocations.}
\label{fig:archiTempl}
\end{figure}

In order to narrow down the system-level architectural templates to focus on, we first make a couple of observations:
\begin{itemize}
\item While some kernels of the simulation loop can be well implemented in HW (PME, PP interactions), there are other parts such as the enforcement of constraints, bonded interactions and integration (double precision) which are better suited for CPUs due to the control flow and flexibility required for algorithmic changes. If we stick to this partitioning, this means that the simulation loop will always involve a complete state exchange (coordinates in, forces out) in each simulation step.
\item Bandwidth from CPU to logic and vice versa is similar for a PCIe card (\SI{15}{\giga\byte\per\second} raw bandwidth) as for an ARM system on a SoC like Stratix 10, Zynq SoC (2$\times$\SI{128}{\bit} plug at, say \SI{400}{\mega\hertz} gives \SI{12.8}{\giga\byte\per\second} of raw bandwidth in total). The only benefit of a SoC will likely be latency, but the embedded CPUs are not as capable as on a server class system. Further, latency is the smaller evil on small single node systems -- it is the bandwidth that is important. I.e., to transfer 2MB of simulation state over \SI{12}{\giga\byte\per\second} \SI{166}{\micro\second}, while link latency is only in the order of a few microseconds.
\item As we have seen in \secref{sec:bench}, a capable server class CPU is required to handle the non-HW-accelerated part of the simulation within the available time budget. An ARM-based SoC system will likely be too slow.
\end{itemize}
Hence, it makes sense to stick to a system architecture template of the form \emph{Xeon Class Server plus N FPGA cards with own interconnect}. This is also aligned with what is currently available in cloud services (AWS F1 for instance), and such a system is depicted in \figref{fig:archiTempl}.

We do not consider IBM Power 8/9 systems in this report as due to their cost and limited availability. Judging from the literature, most people in this space seem to be used to x86 machines, and these machines together with customer grade GPUs are more affordable.
\subsection{Estimated System Components}
As explained before, we assume the architectural template shown in \figref{fig:archiTempl}a). For the FPGA cards, we base the resource and performance estimates on reported values in literature, as explained below.
\subsubsection{Interconnects}
The assumed architectural template consists of two XEON host CPUs that can host a maximum of 4 cards at full x16 PCIe bandwidth, or 8 cards at x8 bandwidth. For the PCIe link efficiency (framing and other overheads), we assume a bandwidth efficiency of 75\% and a latency of \SI{10}{\micro\second}. Further we assume that the FPGA cards are interconnected with a bidirectional ring. This is readily possible due to the high amount of transceivers available on todays high-end FPGAs. In fact, almost every PCIe FPGA card features at least one QSFP port. As explained later, the targeted FPGA platforms either provide two or four QSFP28 cages, allowing for raw ring bandwidths of 200 or \SI{200}{}-\SI{400}{\giga\bit}. The assumed link efficiency is 85\% including framing and packetizing overheads, and a link latency of \SI{0.5}{\micro\second} is assumed.
\subsubsection{PP Interaction Pipelines}
We base our estimates on the PP interaction pipeline by M. Herbordt's group \cite{Chiu2009efficient,Chiu2011efficient} that employs particle filters that generate the neighbor lists on-the-fly. This architecture is well suited for hardware implementations, and does not require as much memory as an implementation with explicit neighbor-lists. The employed arithmetic is a mix between fixed-point and single-precision floating-point, and has been optimized for FPGAs. The employed domain decomposition is an optimized variant of the half shell method. Better methods with less inter-cell communication volume exist (e.g., neutral territory and mid-point methods \cite{Bowers2006}), but these are likely to show their benefits only in the highly scaled regime (which is not the scope of this evaluation). According to the results of Herbordt~et~al., the preferred design employs first-order interpolation with piecewise polynomials, and this design amounts to around 14.5k ALMs and 82 DSP multipliers on a Stratix IV (including filters, LJ and Coulomb datapath, distribution and accumulation logic). We estimate the needed amount of memory separately as described further below.

The PP interaction pipelines can calculate one interaction per cycle. In terms of utilization, the authors reported that it is possible to achieve relatively high percentages in the order of 95\% by properly mapping the particles to the filters and LJ/Coulomb force datapaths. However, since in our evaluation we deal with many more parallel pipelines (several 100 instead of several tens), we assume a slightly reduced utilization of 80\% to account for potential imbalances.

The required RAM resources have been estimated by assuming that each particle position entry requires three \SI{32}{\bit} coordinate entries plus one \SI{32}{\bit} entry holding metadata like atom ID and type (we assume that atom specific LJ interaction values and charge parameterizations can be compiled into ROMs and indexed by these atom type IDs at runtime). The particle force accumulation entries are assumed to comprise three \SI{32}{\bit} values. The estimation for memory usage includes all interpolator lookup tables, atom property ROMs, all position+metadata entries and private accumulation buffers of the PP interaction pipelines.
\begin{table}[!t]
  \caption{Assumed costs for the comparison. Equipment and software packages are amortized over 5 years (straight line), and the electricity is assumed to cost 0.1\$ per kWh. Further, for each on-site solution we double the electricity cost in order to account for cooling.}
 \label{tab:cotsTable}
 \renewcommand{\arraystretch}{0.95}
 \centering
 \small
 \begin{threeparttable}
 \begin{tabularx}{0.48\textwidth}{@{}p{2.85cm}|r|c|l@{}}
 \toprule
  \textbf{Component}          & \textbf{Cost [\$]} & \textbf{[kW]} & \textbf{Details} \\
  \midrule
  Server with 4$\times$PCIe$^\dag$ &  8'000       & 0.45   & Dual E5-2640v4 \\
  Server with 8$\times$PCIe$^\dag$ & 11'000       & 0.45   & Dual E5-2640v4 \\
  P100                        & 6'000        & 0.25   & Pascal Generation\\
  GTX1080Ti                   & 800          & 0.25   & Pascal Generation\\
  TITAN-XP                    & 1'350        & 0.25   & Pascal Generation\\
  V100 SMX2                   & 10'000       & 0.25   & Volta Generation\\
  TITAN-V                     & 3'600        & 0.25   & Volta Generation\\
  VCU1525 (VU9P)              & 4'500        & 0.25   & 200GBit Ring\\
  XUPP3R (VU9P)               & 10'000       & 0.25   & 400GBit Ring\\
  XUPP3R (VU13P)              & 15'000       & 0.25   & 400GBit Ring\\
  DE10-PRO (GX2800)           & 15'000       & 0.25   & 400GBit Ring\\
  \midrule
  p3.2xlarge$^\ddag$          & 3.305        & --     & 1$\times$V100\\
  p3.8xlarge$^\ddag$          & 13.22        & --     & 4$\times$V100\\
  f1.2xlarge$^\ddag$          & 1.815        & --     & 1$\times$XUPP3R (VU9P)$^\ast$\\
  f1.16xlarge$^\ddag$         & 14.52        & --     & 8$\times$XUPP3R (VU9P)$^\ast$\\
  \midrule
  Amber License$^+$           & 20'000/4     & --     & Commercial, Site\\
  \bottomrule
 \end{tabularx}
 \begin{tablenotes}
 \item[$\dag$] Assuming the same dual Xeon configuration of the Sassauna node used for benchmarking.
 \item[$\ddag$] For the AWS instances, the prices are per hour.
 \item[$+$] The Amber license is a site license. We assume here for simplicity that a site consists of 4 nodes to amortize the license cost.
 \item[$\ast$] Or a similar FPGA card that provides up to \SI{400}{\giga\bit} ring interconnection capability.
 \end{tablenotes}
 \end{threeparttable}
\end{table} 
\begin{table*}[!h]
 \caption{Considered FPGA configurations (each listed solution employs 7 PP FPGA cards and 1 PME FPGA card).}
 \label{tab:estimationsTable}
 \newcommand{\al}[2]{\multicolumn{1}{#1}{#2}}
 \renewcommand{\arraystretch}{0.95}
 \centering
 \small
 \begin{threeparttable}
 \begin{tabularx}{0.98\textwidth}{@{}p{4.2cm}|x{2.3cm}|x{2.3cm}|x{2.3cm}|x{2.3cm}|x{2.3cm}@{}}
\toprule
\multicolumn{6}{c}{\textbf{Evaluated Configurations}} \\
\midrule
\textbf{Platform/Board}                    & VCU1525                & XUPP3R                 & XUPP3R                  & DE10 Pro               & f1.x16large$^\ddag$\\
{FPGA}                                     & VU9P                   & VU9P                   & VU13P                   & 1SGX280                & VU9P            \\
\midrule
{Cutoff [\AA]}$^\dag$                      & 12.6                   & 12.3                   & 12.6                    & 12.0                   & 12.0            \\
{PME Grid}$^\dag$                          & 80$^3$                 & 82$^3$                 & 80$^3$                  & 84$^3$                 & 84$^3$          \\
\midrule
{vCPUs}                                    & 40                     & 40                     & 40                      & 40                     & 64              \\
{Core Freq. [MHz]}                         & 400                    & 400                    & 400                     & 700                    & 400             \\
{Ring BW (raw/eff) [GBit]}                 & 200/170$^\mathsection$ & 400/340$^\mathsection$ & 400/340$^\mathsection$  & 400/340$^\mathsection$ & 400/340$^\mathsection$\\
{PCIe BW (raw/eff) [GBit]}                 & 63/54$^+$              & 63/54$^+$              & 63/54$^+$               & 63/54$^+$              & 63/54$^+$       \\
{PP Pipelines / FPGA}                      & 66                     & 66                     & 108                     & 42                     & 58              \\
{Grid Interpolators / FPGA}                & 10                     & 10                     & 18                      & 6                      & 9               \\
{Mixed Radix FFT Units / FPGA}             & 64                     & 64                     & 96                      & 64                     & 64              \\
\midrule
\multicolumn{6}{c}{\textbf{Estimated Resource Utilization}} \\
\midrule
{PP FPGA Logic [kLUT]}                     & \al{r|}{887.9 (75.1\%)}   & \al{r|}{885.8 (74.9\%)}   & \al{r|}{1452.9 (84.1\%)}         & \al{r|}{701.0 (75.1\%)}         & \al{r}{782.1 (85.6\%)}   \\
{PP FPGA DSP}                              & \al{r|}{2706.0 (39.6\%)}  & \al{r|}{2706.0 (39.6\%)}  & \al{r|}{4428.0 (36.0\%)}         & \al{r|}{1722.0 (14.9\%)}        & \al{r}{2378.0 (42.2\%)}  \\
{PP FPGA Memory [MBit]}                    & \al{r|}{78.3 (23.0\%)}    & \al{r|}{78.3 (23.0\%)}    & \al{r|}{124.4 (27.4\%)}          & \al{r|}{46.7 (20.4\%)}          & \al{r}{62.2 (36.3\%)}    \\
\midrule
{PME FPGA Logic [kLUT]}                    & \al{r|}{846.9 (71.6\%)}   & \al{r|}{846.9 (71.6\%)}   & \al{r|}{1428.4 (82.7\%)}         & \al{r|}{711.3 (76.2\%)}         & \al{r}{794.2 (86.9\%)}   \\
{PME FPGA DSPs}                            & \al{r|}{5888.0 (86.1\%)}  & \al{r|}{5888.0 (86.1\%)}  & \al{r|}{9984.0 (81.2\%)}         & \al{r|}{4352.0 (37.8\%)}        & \al{r}{5504.0 (97.6\%)}  \\
{PME FPGA Memory [MBit]}                   & \al{r|}{50.2 (14.7\%)}    & \al{r|}{54.0 (15.8\%)}    & \al{r|}{50.7 (11.2\%)}           & \al{r|}{57.9 (25.3\%)}          & \al{r}{57.9 (33.8\%)}    \\
\midrule
\multicolumn{6}{c}{\textbf{Estimated Runtimes and Performance}} \\
\midrule
{PME + H2D Transfers[\SI{}{\micro\second}]}       & 274.4                  & 250.9                  & 168.1                   & 196.5                  & 269.0           \\
{PP  + H2D Transfers[\SI{}{\micro\second}]}       & 268.4                  & 250.6                  & 168.1                   & 210.4                  & 264.1           \\
{PP/PME + D2H Transfers [\SI{}{\micro\second}]}   & 307.5                  & 284.0                  & 201.2                   & 243.5                  & 302.1           \\
{SW Overheads [\SI{}{\micro\second}]}             & 280.0                  & 280.0                  & 280.0                   & 280.0                  & 175.0           \\
\midrule
\textbf{Total HW+SW [\SI{}{\micro\second}]}       & \textbf{587.5}         & \textbf{564.0}         & \textbf{481.2}          & \textbf{523.5}         & \textbf{477.1}  \\
\textbf{Performance [ns/d]}                       & \textbf{294.1}         & \textbf{306.4}         & \textbf{359.1}          & \textbf{330.1}         & \textbf{362.2}  \\
 \bottomrule
 \end{tabularx}
 \begin{tablenotes}
 \item[$\dag$] We employ a similar load balancing mechanism as GROMACs between PME and PP cards.
 \item[$\ddag$] The maximum available resources are reduced in this case due to the AWS shell infrastructure (see text for more details).
 \item[$+$] We assume a PCIe link efficiency of 75\%, and since we use 8 PCIe cards only half the 16x bandwidth is available per FPGA.
 \item[$\mathsection$] Assuming a link efficiency of 85\% for the ring interconnect.
 \end{tablenotes}
 \end{threeparttable}
\end{table*}

% toPrint = {'%s', name;...
%           '%.1f', cut;...
%           '%.0f', pmeGrid(1);...
%           '%.0f', nCpu;...
%           '%.0f', coreFreq/1e6;...
%           '%.0f', bw_fpgaRing/1e9;...
%           '%.0f', nPipesPerFpga;...
%           '%.0f', nGridInterpPerFpga;...
%           '%.0f', nFftPerFpga;...
%           '%.1f (%.1f%%)', [ppInteractionResourcesFpga(1)/1000, ppInteractionResourcesFpga(1)/fpgaResources(1) * 100];...
%           '%.1f (%.1f%%)', [ppInteractionResourcesFpga(3),      ppInteractionResourcesFpga(3)/fpgaResources(3) * 100];...
%           '%.1f (%.1f%%)', [ppInteractionResourcesFpga(4)/1e6,  ppInteractionResourcesFpga(4)/fpgaResources(4) * 100];...
%           '%.1f (%.1f%%)', [pmeResourcesFpga(1)/1000, pmeResourcesFpga(1)/fpgaResources(1) * 100];...
%           '%.1f (%.1f%%)', [pmeResourcesFpga(3),      pmeResourcesFpga(3)/fpgaResources(3) * 100];...
%           '%.1f (%.1f%%)', [pmeResourcesFpga(4)/1e6,  pmeResourcesFpga(4)/fpgaResources(4) * 100];...
%           '%.1f', times.t_gatherDone/1e-6;...
%           '%.1f', times.t_ppDone/1e-6;...
%           '%.1f', times.t_d2hDone/1e-6;...
%           '%.1f', times.t_hostOverhead/1e-6;...
%           '%.1f', (times.t_hostOverhead+times.t_d2hDone)/1e-6;...
%           '%.1f', perf;...
%           }; 
%
\subsubsection{PME Unit}
Herbordt's group demonstrated as well that it is possible to fit a 3D FFT unit with 64$^3$ grid points onto recent FPGAs \cite{Humphries20133d,Humphries20143d}. The dominant factor here are the 1D FFT macros provided by the FPGA vendors. Hence, we use the complexity of these vendor provided macros to estimate the resources for the 3D FFT part. Since our FFTs sizes are in the order of 84$^3$ grid points, we assume that the resource consumption is similar to 128-point FFT macros, i.e., 5k LUT, 32 DSP multipliers and 16kBit memory on a VUP FPGA\footnote{Note that these implementations likely use a Cooley-Tukey FFT that can only be used for lengths that are powers of two. In practice, a split-radix FFT would be required to handle other FFT lengths.}. The latency corresponds to the amount of samples to be calculated.

For the particle-to-grid and grid-to-particle interpolators, we use the results reported in \cite{Sanaullah2016fpga}, where an optimized single-cycle datapath is designed and implemented. One such unit requires 51k ALMs 192 DSP blocks ($=2\times192$ DSP multipliers) on a Stratix V FPGA. Further, we assume that this design can be optimized such that the same interpolators can be used for both interpolation directions, as well as the PME solution step in the frequency domain that entails a point-wise multiplication with pre-computed constants.
\subsubsection{Resource Allocation}
In terms of allocation of PME and PP units to accelerator boards it seems natural to uniformly distribute them as illustrated in \figref{fig:archiTempl}b). The advantage of this allocation is scalability of compute resources. However, the communication pattern of the two 3D FFT passes in PME leads to high communication volume between these distributed PME units, and hence it has been found that a \emph{contraction} in the PME calculation can lead to better performance \cite{Sheng2017a}. E.g., for highly parallel scenarios GROMACS supports the use of fewer separate MPI ranks \cite{Kutzner2015}. Further, an additional optimization for single node systems has been added in the newest GROMACS release where the PME calculation can be allocated to one GPU. AMBER moved to single GPU implementations already a while ago to solve this communication issue. Hence, we do not consider uniform allocation, but the non-uniform allocation shown in \figref{fig:archiTempl}c).
\subsubsection{Schedule}
The computation schedule follows a relatively simple repetitive pattern. In each iteration, all MPI processes on the host push the particle coordinates and meta-information down to the FPGA card using bulk DMA transfers, thereby utilizing the full effectively available PCIe bandwidth. Computation on the FPGA side can be largely overlapped with the PCIe transfers, since we can start computing PP interactions already with a small part of the simulation volume. Further, the coordinates and meta data can be gathered and transferred to the PME card on-the-fly (via the ring interconnect), and the grid interpolators can start with particle-to-grid interpolation. As soon as the complete simulation volume has been transferred, the PP cards exchange the overlap regions required for PP interactions via the ring. Once the inverse 3D FFT is complete, the grid-to-particle interpolation can be started and overlapped with the scatter operation that transfers the forces back to the originating card. The forces are then accumulated within the PP interaction force buffers, and once all non-bonded forces have been calculated, the results are copied back to the host.
\subsection{FPGA Targets and Hardware Costs}
In order to achieve the required performance, large datacenter grade FPGAs are required. In this report we target the Xilinx Virtex UltraScale+ series, as well as the Intel Stratix 10 series, since at the time of writing, these represent the best FPGA technology that is available (or soon will be). On the Xilinx side, we identified the VU9P device as an ideal target as it is currently widely used and stable in production (this device is also available in the AWS F1 instances). The roughly 50\% larger VU13P device will soon reach stable production, as well, and can be seen as the natural successor of the VU9P device in the forthcoming years. On the Intel side, the only FPGA that can currently match the Virtex UltraScale+ devices is the upcoming Stratix 10 series (the Arria 10 devices are too small). In particular, it seems that the 1SGX280 device will be the equivalent of the VU9P in terms of adoption (several development boards feature this device). However, we found that the availability of these devices is not yet guaranteed (especially for the H-Tile devices with fast transceivers), and we can expect that stable products are likely not to going to be introduced before next year.

In order to compare system level costs, we assume the prices and power consumptions\footnote{For more information on the FPGA boards, see also:\\
\url{http://www.hitechglobal.com/Boards/UltraScale+_X9QSFP28.htm}, \\
\url{https://www.xilinx.com/products/boards-and-kits/vcu1525-p.html}, \\
\url{https://www.bittware.com/fpga/xilinx/boards/xupp3r/}, \\
\url{https://www.altera.com/solutions/partners/partner-profile/terasic-inc-/board/terasic-stratix-10-de10-pro-fpga-development-kit.html}} listed in \tabref{tab:cotsTable}.
For simplicity, we lump the costs for the optical modules together with the corresponding FPGA boards (roughly 200\$ per QSFP28 slot). Equipment and software packages are amortized over 5 years (straight line), and the electricity is assumed to cost 0.1\$ per kWh. Further, for each on-site solution we double the electricity cost in order to account for cooling. No sales margin is added to the FPGA solutions in this comparison, but in a commercial setting this has to be accounted for as well.
\subsection{Evaluated Configurations and Assumptions}
We calculate our estimates for a system with $N=8$ FPGA cards in the system (as discussed earlier, the PP interaction pipelines are allocated on 7 FPGAs, while 1 FPGA card is used for PME). These configurations are listed in \tabref{tab:estimationsTable}. The amount of units (e.g. FFTs) listed in that table is per FPGA instance. We note that on these modern FPGAs, logic resources are the ones that are going to be critical. In order to allow for enough headroom for additional infrastructure such as, e.g., Aurora and PCIe PHYs we target a LUT/ALM usage of 75\% on the VU9P and 1SGX280 devices (both offer a similar amount of logic resources). On the larger VU13P devices and AWS F1 instance, we target a higher utilization in the order of 85\%. This is possible since the VU13P offers around 50\% more logic resources than the VU9P, and on the AWS instance, the infrastructure is already included in the AWS F1 Shell that wraps the user logic\footnote{AWS uses partial reconfiguration, and the logic resources available in the user partition amounts to 914k, 5640 DSPs, 3360 BRAMs and 400 URAMs, which corresponds to 77\%, 82\%, 78\% and 42\% of the resources available on the VU9P.}.

From test syntheses of an HDL design optimized for FPGAs (NTX cores from \cite{Schuiki2018}) we found that operating frequencies up to 400MHz and 700MHz should be achievable on the Virtex UltraScale+ and Stratix 10 devices, respectively. From these test syntheses we also calculated logic conversion factors for the LUTs and ALMs to derate the numbers reported for the 3D FFT and PP cores in literature (shown in \tabref{tab:convFactTable}).

\begin{table}[!t]
 \caption{Derating factors for different FPGA generations. The higher \# ALMs on the Stratix 10 device is likely due the high operating frequency targeted.}
 \label{tab:convFactTable}
 \renewcommand{\arraystretch}{0.95}
 \centering
 \small
 \begin{threeparttable}
 \begin{tabularx}{0.48\textwidth}{@{}p{3.3cm}|cc@{}}
 \toprule
 \textbf{Source / Target} & VUP [LUTs] & Stratix 10 [ALMs] \\
 \midrule
 Stratix IV [ALMs]        & 0.95       & 1.18 \\
 Stratix V  [ALMs]        & 1.04       & 1.29 \\
 \bottomrule
 \end{tabularx}
 \begin{tablenotes}
 \end{tablenotes}
 \end{threeparttable}
\end{table} 
For the estimations in this section, we consider the 91k problem (2LEM) of the previous section, and assume a software overhead of \SI{280}{\micro\second} on 40 logical cores that can not be overlapped with the force computations (this amounts to 25\% of the overall walltime of a single step, see previous section on benchmarks). Note that we adjust the PME grid resolution and PP interaction cutoffs to balance the load between the PP and PME cards at similar accuracy. This is analogous to the PME load balancing procedure performed in GROMACS simulation runs \cite{Kutzner2015best}.

We also note that we do not explicitly account for potential and virial calculations here as these are only carried out every 100 steps in the considered benchmarks. The support for these calculations can be added to the hardware either by extending the force pipelines (this leads to a small increase in DSP slices which are still abundantly available) or by reusing existing interpolator infrastructure and repeated evaluation, leading to a decrease in performance of around 1\%.
\subsection{Results}
The estimations for resources and timings are shown in \tabref{tab:estimationsTable} at the bottom. The highest performance is achieved by the VU13P and f1.16xlarge designs. In the first case this is due to the high amount of logic resources on the VU13P and in the second case, more CPU cores translate into lower software overheads. In all cases we note that the non-hideable software overheads (domain decomposition, integration, constraints, etc.) are either in similar in magnitude than the accelerated PP and PME portions.

When comparing the VCU1525 and XUPP3R solutions with VU9P FPGAs, we can observe that the faster ring interconnect available with the more expensive XUPP3R does lead to a small improvement in speed, but this is likely not worth the price increase of around 2$\times$ in case of the VU9P. However, for the larger/faster VU13P and 1SGX280 FPGAs, the faster ring interconnect is desirable in order to match the bandwidth with the increased throughput.

In order to better compare different solutions, we cast these results into a performance (ns/d) versus cost (\$/h) and add different operating points of GPU-accelerated solutions. The performance values for FPGA versions with less than eight cards have been derated from the 8 card solutions (assuming that one of the cards now contains both the complete PME unit and some PP interaction pipelines). The performance values for the GPU solutions employing 1-4$\times$GTX1080Ti and 1 V100 (on AWS) have been measured (see previous section on benchmarks). The remaining operating points have been estimated using numbers from existing benchmarks\footnote{See \url{http://ambermd.org/gpus/benchmarks.htm}}.

The blue dots are all for GROMACS 2018, the green ones for AMBER 16/17 and the orange ones for FPGA solutions based Virtex UltraScale+ and Stratix 10 FPGAs. On the right side of the plot we have the AWS instances, and on the left the on-prem versions. The closer solutions are to the upper left corner of the plot, the better, and the diagonal lines represent same performance/cost. The red dot in the upper left corner shows the desirable performance/cost of an ideal solution that domain experts would consider commercially feasible.
\begin{figure}[!t]
\centering
\includegraphics[width=0.48\textwidth]{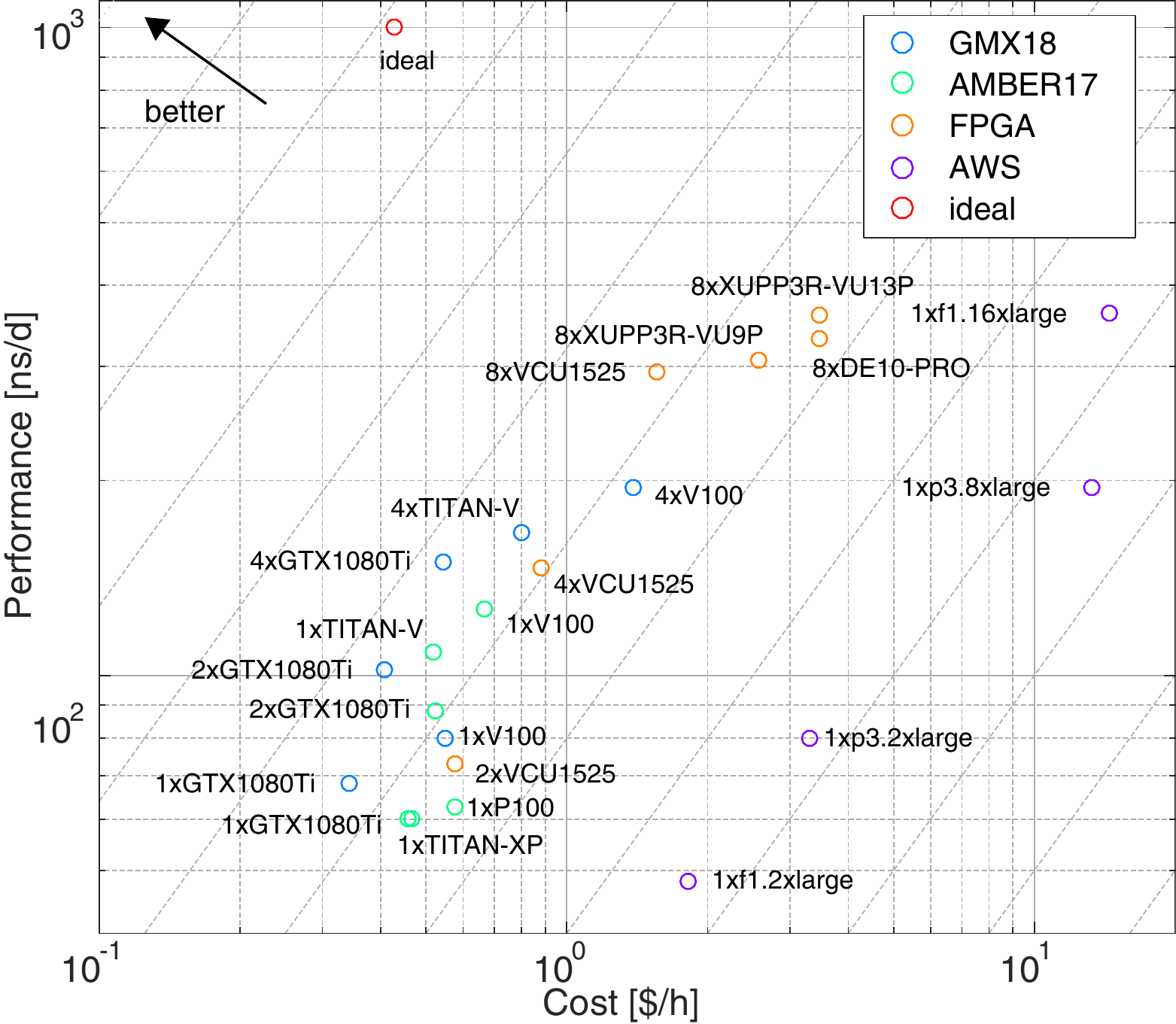}
\caption{Comparison of the performance vs cost tradeoff of different FPGA and GPU-based solutions on the 2LEM-91k benchmark.}
\label{fig:costScatterPlot}
\end{figure}
\subsection{Discussion}
As can be observed in \figref{fig:costScatterPlot}, systems employing consumer grade GTX1080Ti cards are clearly at the forefront in terms of cost efficiency (slanted lines represent equal performance/cost). And it should be noted that this efficiency is bound to increase, since there seems to be a trend in moving more computations or even the complete simulation loop onto the GPUs and have them communicate in a peer-to-peer fashion. AMBER already supports this, allowing to assemble very cost efficient desktop systems, since in that case no expensive high-core-count CPUs are required anymore. This is not reflected in the plot above yet, as these costs have been calculated assuming a dual XEON (20 core) server.

We can also see that with FPGA solutions based on UltraScale+ and Stratix 10 devices there is not much to be gained with respect to the GPU solutions. I.e., it is possible to achieve speedups in the order of around 1.5$\times$ to 2$\times$, but the performance/cost ratio is similar to GPU solutions. This is a combination of two key factors: first, FPGAs prices are in the range of datacenter GPUs, which makes it difficult to compete with consumer grade GPUs that offer very attractive single-precision FLOPS/\$ ratios. And second, the amount of remaining work that can not be overlapped with non-bonded force computations starts to become dominant, thereby leading to a saturation of the achievable speedup. For instance, in the case of a system with 8 VU13P FPGAs, our estimates indicate that the PP and PME calculations take less time than the remaining non-hideable parts in software (\SI{200}{\micro\second} versus \SI{280}{\micro\second}). Hence, we see that even a 4$\times$ speedup of the calculation of the PP interactions and PME only leads to an application performance improvement of only 2$\times$.
\subsubsection{Commercial Feasibility}
We have been in contact with domain experts and according to them, a new accelerator solution based on a different technology than GPUs should offer at least \SI{1}{\micro\second} of simulation performance at the cost of one high-end GPU in order to be perceived as a viable alternative (this \emph{ideal} solution is indicated with a red dot in \figref{fig:costScatterPlot}). Considering our results and this desirable target, it becomes evident that a FPGA co-processor solution will likely not be commercially successful since a mere 2$\times$ improvement in speed at similar cost efficiency does not represent a compelling value proposition for potential users (we have to keep in mind here that an FPGA solution will likely be less flexible in terms of functionality than a GPU-based one). Instead, MD users will likely just run several simulations in parallel at somewhat lower speed and better cost efficiency, e.g., to improve sampling statistics or to screen multiple chemical compounds in parallel.
\subsubsection{A Note on Algorithmic Improvements}
We see that when targeting non-bonded interactions, architectural specialization and current FPGA technology alone will likely not be enough to get the needed improvement in speed (with respect to GPU solutions), and some algorithmic innovation will be needed as well (that is, unless FPGA prices drop over 10$\times$ the coming years, which is not very probable). The following points should however be kept in mind when doing algorithmic work in this field:
\begin{itemize}
\item This is a mathematical field, and people would like to have error guarantees and bounds. Despite the fact that it is very challenging to come up with an algorithm that is better than SOA, it takes a long time until such a new method is established and accepted by the community. For example, variations of FMM and multi-level summation methods (MSM) \cite{Hardy2015msm,Hardy2016multilevel} have been in development for many years now, and they still not used on a regular basis for production runs -- even if they may have benefits in some operating regions. One of the reasons for this slow development and adoption is that it is difficult to find and introduce optimizations/approximations that do not introduce assumptions that violate physical laws and eventually lead to erratic behavior. Further, it is difficult to verify the correctness of a certain approach or implementation.
\item It is likely that GPU-based solutions will benefit from algorithmic improvements, too (e.g., improvements of the integration and constraints steps), and the implementation turn-around time is shorter for GPUs than for FPGAs.
\item Machine-learning approaches are well suited for a certain class of ab-initio potential evaluations (AIMD), and they have been shown to give quite impressive speedups in that case. However, in pure classical MD, the force fields already have a very simple and efficient functional form (basically polynomial expressions for classical mechanics), and contain far fewer parameters than, e.g., the ANI-1 NN potential \cite{Smith2017}. It is currently unclear how such an NN-based approach could be used to accelerate potential expressions in classical MD simulations.
\end{itemize}
\subsubsection{Other Hardware Acceleration Opportunities}
ASIC integration of the non-bonded interaction pipelines could be a way of improving the performance with respect to FPGA-based solutions, but in the end this approach suffers the same shortcomings as the FPGA co-processor solution, and in addition the market does not seem to be big enough to justify the development costs. Moving to fully integrated solutions in a similar fashion as this has been done in the Anton systems can also lead to higher performance, but the design effort (and involved risks) are quite large and hence difficult to justify. So far, Anton-1/2 SoCs have been the only successful chips built in this fashion. MDGRAPE-4, which is the only other SoC-based system, seems to be fizzled out as there has been no update on the project for 4 years (the last publication \cite{Ohmura2014} is from 2014). Another fact that has to be kept in mind is that there are several patent applications and granted patents \cite{bowers2012determining,bowers2012zonal,shaw2017parallel1,shaw2017parallel2} protecting features of the Anton-type SoCs, which could complicate commercial exploitation of such a solution.

\begin{figure}[t]
\centering
\includegraphics[width=0.45\textwidth]{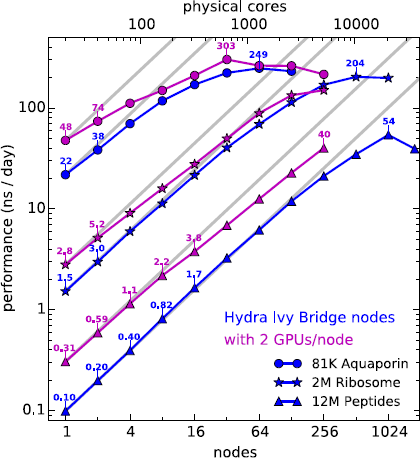}
\caption{GROMACS scaling on HYDRA (Max Planck Computing Centre), reproduced from \cite{Kutzner2015}.}
\label{fig:hydraScaling}
\end{figure}

Considering the difficulties mentioned above, it becomes clear that different approaches for leveraging FPGAs should be considered. It could make sense to turn towards scaled datacenter solutions and look into hybrid solutions that leverage FPGAs as \emph{near-network accelerators} (similar as this has been done in Catapult-I/II \cite{putnam2014reconfigurable,caulfield2016cloud}, or applications such as networking filtering, high-frequency trading, etc). I.e., we could use FPGAs to complement multi-node GPU systems in order to improve the scalability of such systems. Consider for instance the scaling behavior of GPU-accelerated GROMACS runs on the Hydra supercomputer (Max Planck Computing Centre, 20-core Ivy Bridge nodes with 2xK20X and FDR-14) in \figref{fig:hydraScaling}.
We can observe the well-known hard-scaling issue for typical problem sizes (top 2 curves, 81k atoms). Problems with several millions of atoms (lower 4 curves) are often used for benchmarking, but they do not represent common everyday problems. What is interesting to note is that GPU accelerated problems with <100k atoms often reach their 50\% scaling limit very quickly at around 8 nodes - and this is something that can be observed on other clusters, too (see \cite{Kutzner2015best}, for example). From what can be read in literature this scaling bottleneck is mainly due to PME and global communication phases.

So it is likely that a network accelerator could be used to ameliorate this scaling bottleneck by bypassing the standard InfiniBand interconnects and the MPI software stack, and by providing a dedicated secondary network with functional capabilities tailored towards PME and global communication (e.g. for energies and virials) and even integration/constraints handling. Possible arrangements could be 1 FPGA network card per node and a 2D or 3D torus. To this end, initial studies on 3D FFTs for molecular dynamics on FPGA clouds by Herbordt,~et~al.~\cite{Sheng2017a} show promising results (also considering phased contraction where certain parts of the computation are carried out on a subset of all nodes to improve communication patterns). Another arrangement that is similar to phased contraction and suitable for small-scale clusters would be a star arrangement, where all nodes have a connection to a single external FPGA box that performs PME and global reductions/broadcasts. As we have seen with the recent GROMACS update, solving PME on one device alone can be beneficial since this improves the communication volume.
\section{Conclusions}
In this report, we benchmarked two widely used GPU-accelerated MD packages using typical MD model problems, and compare them with estimates of an FPGA-based solution in terms of performance and cost. Our results show that while FPGAs can indeed lead to higher performance than GPUs, the overall application-level speedup remains in the order of 2$\times$. Considering the price for GPU and FPGA solutions, we observe that GPU-based solutions provide the better performance/cost tradeoff, and hence pure FPGA-based solutions are likely not commercially viable. However, we also note that scaled systems could potentially benefit from a hybrid composition, where GPUs are used for compute intensive parts and FPGAs for in-network acceleration of latency and communication sensitive tasks.
\bibliographystyle{IEEEtran}
\bibliography{ms}
\end{document}